\documentclass[showpacs,amsmath,amssymb,twocolumn,superscriptaddress,notitlepage,preprintnumbers,pra]{revtex4-1}

\usepackage{graphicx} 
\usepackage{amsmath,amssymb,amsthm,mathrsfs,amsfonts,dsfont}
\usepackage{braket}
\usepackage{bm}
\usepackage{enumerate}
\usepackage{color,xcolor}
\usepackage{qcircuit}
\usepackage{algorithm}
\usepackage{algpseudocode}
\usepackage{comment}
\usepackage[normalem]{ulem}

\usepackage{varioref}
\labelformat{equation}{#1}

\labelformat{section}{#1}

\labelformat{figure}{#1}

\labelformat{proposition}{#1}

\labelformat{lemma}{#1}

\labelformat{theorem}{#1}

\labelformat{observation}{#1}

\labelformat{definition}{#1}

\labelformat{problem}{#1}

\labelformat{algorithm}{#1}

\labelformat{corollary}{#1}

\labelformat{statement}{#1}

\usepackage{hyperref}
\hypersetup{colorlinks=true, linkcolor=blue, citecolor=blue, urlcolor=black }

\begin{document}

\title{Shaping causality: programmable nonlocal signal generation in long-range spin systems}

\author{Shreyas Sadugol}
\email{ssadugol@tulane.edu}
\affiliation{Department of Physics and Engineering Physics, Tulane University, New Orleans, USA}

\author{Giuseppe Luca Celardo}
\email{giuseppeluca.celardo@unifi.it}
\affiliation{Department of Physics and Astronomy and CSDC, University of Florence, Florence, Italy}
\affiliation{European Laboratory for Non-Linear Spectroscopy (LENS), University of Florence, Florence, Italy, 50019 Sesto Fiorentino, Italy}
\affiliation{Istituto Nazionale di Fisica Nucleare (INFN), Sezione di Firenze, 50019 Sesto Fiorentino, Italy}
\author{Fausto Borgonovi}
\email{fausto.borgonovi@unicatt.it}
\affiliation{Department of Mathematics and Physics and ILAMP, Catholic University of the Sacred Heart, Brescia, Italy}
\affiliation{INFN, Sezione di Milano, Italy}

\author{Lev Kaplan}
\email{lkaplan@tulane.edu}
\affiliation{Department of Physics and Engineering Physics, Tulane University, New Orleans, USA}
\date{\today}

\begin{abstract}
Understanding how information spreads in non-relativistic many-body systems is a central issue for quantum information processing~\cite{Jurcevic2014, Monroe2021}. While short-range interactions confine information within a local light cone~\cite{LiebRobinson1972}, long-range interactions typically lead to uncontrolled nonlocal spread across the entire system~\cite{Tran_2021, Richerme2014}. Here, we demonstrate that this apparent dichotomy is not fundamental and that nonlocality in systems with long-range interactions can be deterministically controlled. By mapping spin dynamics to a hard-core boson chain, we identify a regime 
in which the causal space-time landscape can be precisely shaped. 
We show that placing spin excitations in a polarized background allows a local perturbation to trigger nonlocal signals exactly at the positions of these excitations. These pre-selected sites act as seeds for new, effective light cones, allowing information to bypass the bulk and re-emerge at distant, programmable locations. This mechanism avoids uncontrollable global nonlocality while circumventing the speed limits associated with local transport. By engineering these nonlocal communication channels, our findings offer a versatile framework for information distribution relevant to quantum memories, error correction, and programmable platforms such as trapped ions~\cite{Monroe2021,lanyon2011universal,Korenblit2012}.


\end{abstract}

\maketitle

\let\oldaddcontentsline\addcontentsline
\renewcommand{\addcontentsline}[3]{}


\begin{figure*}[htbp!]
    \centering
    \includegraphics[width=\linewidth]{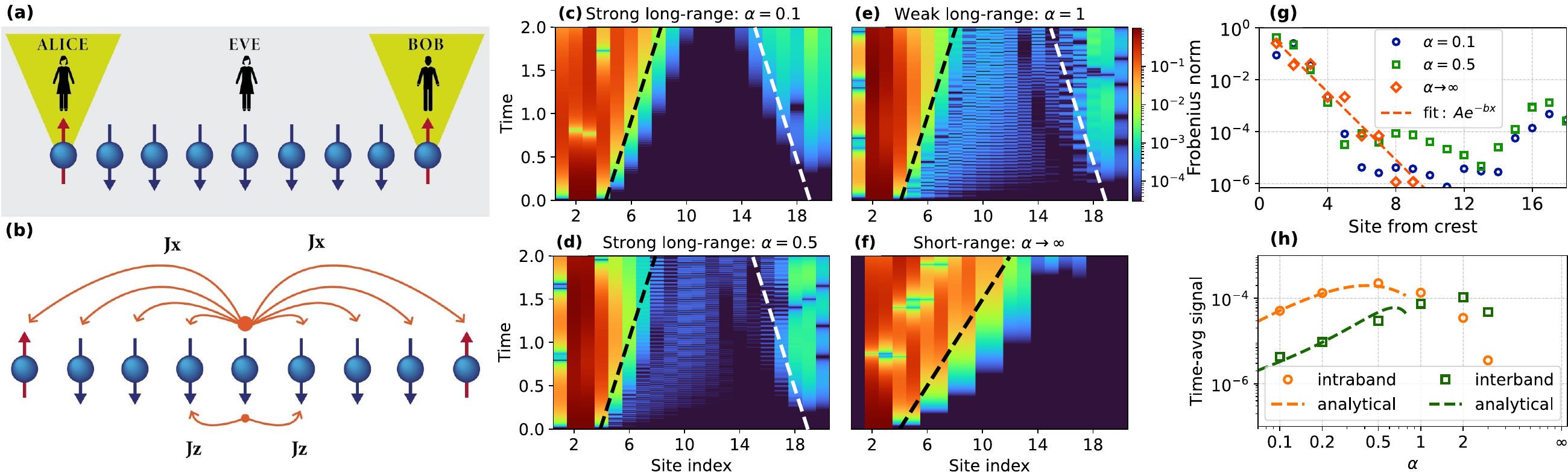}
    \caption{
     Illustration of: \textbf{(a)} nonlocal signaling,  \textbf{(b)}  the model. \textbf{(c-h)} Emergence of nonlocal light cones for  $N=20$, $J_x=5$, and $J_z=1$.
    \textbf{(c–f)} Density plots of the Frobenius norm $\|\Delta \rho_n(t)\|_F$ showing light-cone dynamics under the Hamiltonian \eqref{HAM} for initial states prepared in band~2, $(2,19;\,3,19)$ (notation defined in Eq.~\eqref{pairnotation}). Dashed lines (black and white for contrast) indicate the analytically predicted light-cone velocities: $2J_z$ for interaction exponent $\alpha \le1$ and $4J_z$ for $\alpha\to\infty$
    (see Supplementary Information). A nonlocal light cone is clearly visible in the long-range regime ($\alpha=0.1, 0.5, 1$) and disappears in the nearest-neighbor coupling limit, $\alpha\to\infty$, \textbf{(f)}.
    \textbf{(g)} Outside-the-cone profiles at $t=0.7$, obtained by locating the crest of the outward-propagating wave and plotting the decay of $\|\Delta \rho_n(t)\|_F$ beyond the wave crest. All profiles decay exponentially (consistent with the Lieb–Robinson bound), with the long-range cases exhibiting an even steeper decay than the nearest-neighbor case whose profile is fitted by an exponential (orange dashed line with $A=1.3$, $b=1.5$). \textbf{(h)} Time-averaged $(t\in[0,0.5])$ signals at sites 19 and 10 showing respectively the intraband and interband signals as a function of interaction exponent $\alpha$. The dashed curves show the predicted analytical small-$\alpha$ scaling in Eqs.~\eqref{Interband_main} and \eqref{Intraband_main}, up to an overall constant fitting factor.}   
    \label{N20LFIMb2VaryalphaJx5}
\end{figure*}

\textit{\textbf{Introduction---}}
\label{sec:introduction}
Although non-relativistic quantum systems are not fundamentally constrained to exhibit a finite speed for information propagation, information in many-body systems with short-range interactions nevertheless spreads in a locally bounded manner, as formalized by the Lieb–Robinson bounds~\cite{LiebRobinson1972}.
These bounds dictate that the spread of information occurs at most ballistically within an effective light cone, with a velocity determined by the coupling strength, and with leakage of information outside the light cone being at least exponentially suppressed. Thus, locality
limits information propagation and shapes our understanding of entanglement and correlation growth, thermalization, and causality~\cite{LiebRobinson1972,Hastings2006,sugimoto2022eigenstate,kastner2011diverging, Chen2023SpeedLimits}.

Many programmable quantum platforms~\cite{Defenu_2023} naturally realize long-range couplings. These include trapped ions~\cite{Jurcevic2014, Monroe2021, blatt2012quantum, lanyon2011universal, Richerme2014, noel2022measurement}, Rydberg atom arrays~\cite{Browaeys2020, Bernien2017, Bluvstein2022}, dipolar molecules~\cite{Yan2013, Schindewolf2022}, and optical cavity QED systems~\cite{Baumann2010, Landig2016, Vaidya2018, Ritsch2013}. Such interactions can enable faster-than-ballistic spread of information, suggesting a breakdown of locality~\cite{Tran_2021, Richerme2014, Jurcevic2014, eisert2013breakdown, metivier2014spreading, Halati_2025}.

However, simply increasing the interaction range often leads to uncontrolled spread of information~\cite{Cevolani_2016, Gong_2014, Kuwahara_2020, Luitz_2019, Feig_2015}. To control this behavior and engineer useful quantum channels, one requires a mechanism that suppresses generic nonlocal transport while enabling targeted signaling. We demonstrate that this can be achieved by exploiting the interplay between long- and short-range interactions.

Here, we show that for a class of long-range 1D spin chains, the causal space-time landscape can be precisely shaped. We demonstrate the existence of a parameter regime (large system size or strong long-range coupling) where the onset of nonlocality is fully programmable. To illustrate this mechanism, we consider a transmission protocol involving three parties: a sender (Alice), an intermediate observer (Eve), and a distant receiver (Bob) [Fig.~\ref{N20LFIMb2VaryalphaJx5}(a)].
While short-range interactions require Alice’s signal to traverse every intermediate site to reach Bob, our long-range architecture breaks this constraint. By tuning the initial state, Alice can trigger a nonlocal signal that ``jumps'' over Eve entirely, appearing at Bob's site while leaving Eve undisturbed. Once it emerges at the target site, the signal reverts to local propagation within a linear light cone. This creates a direct, nonlocal communication channel that bypasses the geometric bulk of the chain.

Crucially, this nonlocal signaling is deterministic and scalable: 
by placing additional background excitations at other positions (e.g., another observer, Charlie), Alice can broadcast the nonlocal signal to multiple, arbitrarily chosen destinations simultaneously, [see, e.g., Fig.~\ref{Prjb3}(a-c)]. Crucially, this mechanism extends to higher dimensions, see Supplementary Information.\\

\textit{\textbf{The model---}}
\label{sec:model}
We demonstrate this physics using 1D long-range interacting spin-$1/2$ systems with open boundary conditions, Fig.~\ref{N20LFIMb2VaryalphaJx5}(b), described by the Hamiltonian:
\begin{equation}
\label{HAM}
\hat H = \sum_{j=1}^{N-1} J_z\, \sigma_j^z \sigma_{j+1}^z
+\sum_{i<j} \frac{J_x}{|j - i|^\alpha} \,\sigma_i^x \sigma_j^x \,,
\end{equation}
where $J_z,J_x>0$ (setting $J_z=1$ in the following for convenience), and $\alpha$ controls the interaction range, interpolating between infinite-range ($\alpha=0$) and nearest-neighbor ($\alpha \to \infty$) limits. Note that applying Kac rescaling~\cite{Botzung_2021, Mori_2012, kastner_2025} for thermodynamic consistency does not affect our main results (see Supplementary Information).

\begin{figure*}[htbp!]
    \includegraphics[width=\linewidth]{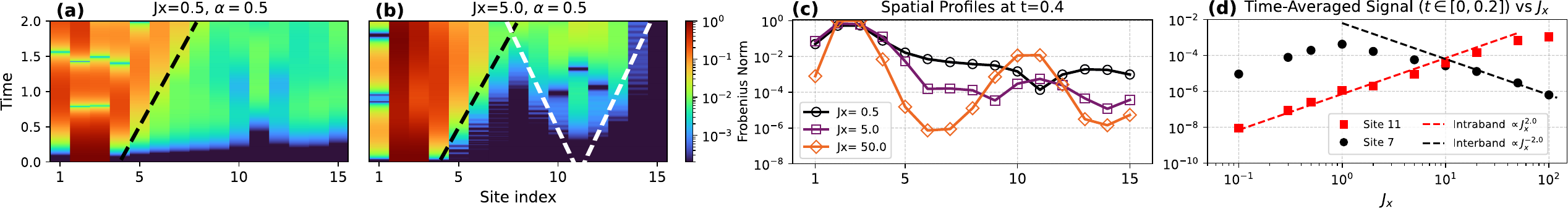}
    \caption{\textbf{(a,b)} Density plots of the Frobenius norm for $N=15$, $\alpha=0.5$ showing light-cone dynamics
    for initial states prepared in band~2, $(2,11;\,3,11)$ (notation defined in Eq.~\eqref{pairnotation}). Dashed lines (black and white for contrast) indicate the analytically predicted light-cone velocities (see Supplementary Information).  \textbf{(c)} Spatial profiles of the light cone at a fixed time $t=0.4$. Increasing $J_x$ suppresses interband contributions, leading to the clear emergence of a well-defined nonlocal light cone. 
    \textbf{(d)} Scaling behavior of the time-averaged signal. At site 11 (red squares), the time-averaged signal scales as $J_x^{2}$, in conformance with the analytical intraband scaling formula, Eq.~\eqref{Intraband_main}. Conversely, at site 7 (black circles), the signal exhibits an asymptotic $J_x^{-2}$ scaling in the large-bandgap limit, see Eq.~\eqref{Interband_main} and Supplementary Information for more details.}
    \label{NatureFig2}
\end{figure*}

To quantify the information spread, 
we use the Frobenius-norm distance between the reduced density matrices $\rho^{(1)}_{n}(t)$ and $\rho^{(2)}_{n}(t)$, respectively, of  two evolved initial states at each site $n$:
\begin{equation}
\label{eq:fnorm}
\|\Delta \rho_n(t)\|_F = \sqrt{\text{Tr}(\Delta\rho_n(t) \Delta\rho_n^\dagger(t))}\,,
\end{equation}
where \( \Delta\rho_n(t) = \rho^{(1)}_{n}(t) - \rho^{(2)}_{n}(t) \). We normalize this value in the interval $[0,1]$, see Methods.
Unlike two-point correlators, OTOCs, or magnetization, which test specific observables in specific orientations, the Frobenius norm in Eq.~\eqref{eq:fnorm} captures the maximum possible change in any measurement outcome, making it sensitive to coherent differences that other probes can miss.  

Information propagation can also be evaluated using the site-resolved $x$-magnetization $\left|\frac{\bra{\psi_{1}}\sigma^x_n\ket{\psi_{1}} - \bra{\psi_{2}}\sigma^x_n\ket{\psi_{2}}}{2}\right|$ or the  two-point correlator $|\langle \sigma_2^x(t)\sigma_j^x(t)\rangle
    - \langle \sigma_2^x(t)\rangle \langle \sigma_j^x(t)\rangle|$, as has been done in ion trap experiments~\cite{Richerme2014}. In the latter case, only a single initial state is required. Both figures of merit  can be used to unveil the same physics described by the Frobenius norm, see  Supplementary Information.

Specifically, to compute the normalized  Frobenius norm, we compare the time evolution of two nearby initial states, $\ket{\psi_1}$ and $\ket{\psi_2}$. As an illustrative example, consider a chain of $N=20$ spins and 

\begin{align}
\label{eq:psi12}
\ket{\psi_1} &= \ket{\downarrow_1\uparrow_2\downarrow_3\downarrow_4\cdots\uparrow_{19}\downarrow_{20}}, \nonumber\\
\ket{\psi_2} &= \ket{\downarrow_1\downarrow_2\uparrow_3\downarrow_4\cdots\uparrow_{19}\downarrow_{20}}\,,
\end{align}
which differ by a single-site shift of one spin-up excitation  (site $2 \leftrightarrow 3$, with the second excitation fixed at site $19$). All spins are specified in the $\sigma^x$ basis, i.e., along the direction defined by the long-range interaction.
We denote such pairs of initial states compactly as
\begin{equation}
(i_1,i_2;\, j_1,j_2),
\label{pairnotation}
\end{equation}
where $(i_1,i_2)$ and $(j_1,j_2)$ specify the excitation positions in $\ket{\psi_1}$ and $\ket{\psi_2}$, respectively; here $(2,19;\,3,19)$.

Physically, this setup corresponds to a differential signaling protocol: Alice encodes information by choosing between two different local excitations at the source (site 2 vs. 3). Bob retrieves this information by monitoring how the system's evolution (state $\ket{\psi_2}$) deviates from a known calibration baseline (state $\ket{\psi_1}$).

An example of nonlocal light cones emerging from  the interplay of long- and short-range interactions for strong long-range coupling $J_x$ is shown in Fig.~\ref{N20LFIMb2VaryalphaJx5}(c-e).
We compute the Frobenius norm by evolving the states in Eq.~\eqref{eq:psi12} with the Hamiltonian in Eq.~\eqref{HAM} 
for different values of the interaction exponent $\alpha$.
For long-range interactions  (c-e), a second light cone erupts at the target site 19, despite the two initial states being identical in that region. As a comparison, for the same initial conditions with nearest-neighbor interactions ($\alpha=\infty$), a single light cone emerges from the source perturbation,  Fig.~\ref{N20LFIMb2VaryalphaJx5}(f).
The information leaking outside the nonlocal, receiver-centered light cone (site 19) decays faster than exponentially, matches the behavior observed for the light cone originating at the source, see Fig.~\ref{N20LFIMb2VaryalphaJx5}(g) and Supplementary Information for more details. The intensity of the signal both within the nonlocal light cone and in the intermediate region between the source and receiver depends on several parameters. In particular, their dependence on the interaction exponent $\alpha$ is shown in Fig.~\ref{N20LFIMb2VaryalphaJx5}(h). 
As $\alpha$ increases, both the signal inside the nonlocal light cone and between the two light cones increase at first, while for $\alpha \gg 1$ both signals decrease and only the light cone at the source remains. \\

\textit{\textbf{Band structure and mapping to a hard-core boson model---}} 
\label{sec:mapping}
The emergence of these programmable nonlocal light cones can be rigorously understood by analyzing the band structure of the excitation spectrum and by mapping the spin Hamiltonian to a hard-core boson chain, as detailed below.

Let us consider the band structure of the spectrum of long-range interacting systems. For $\alpha = 0$, the long-range term depends only on the total $x$-magnetization $M_x = \tfrac12 \sum_{j=1}^N \sigma_j^x$, so the spectrum organizes into well-separated degenerate “bands” (or excitation manifolds) labeled by the number $b=0,1,2,...$ of up spins along $x$. The band energies $E_b$ are quadratic in $b$ with large interband gaps $\Delta E_b$ that strongly suppress transitions between them. Explicitly, one has:
\begin{align}
\label{bandgap}
    E_b &=  \frac{J_x}{2}(N - 2b)^2 - \frac{J_x N}{2}, \\
\Delta E_b &= E_b-E_{b-1} = -2 J_x \left(N-2b+1 \right) \,.
\end{align}
The $J_z$ nearest-neighbor term couples only bands $b \leftrightarrow b \pm 2$.

For $0<\alpha<1$, the many-body bands remain, but intraband degeneracies lift. For band 1, the band width $\delta E_1$ is set by the energy difference between excitations at the chain center and edge, which for small $\alpha$ behaves as

\begin{equation}
    \frac{\delta E_1}{|\Delta E_{1}|} 
\;\sim\; \frac{2 J_x (\ln 2)\alpha \, N^{\,1-\alpha}}
{2 J_x \, N^{\,1-\alpha}}
\;=\; (\ln 2)\alpha \,.
\end{equation}
As $\alpha \to 1^-$, the ratio becomes $\mathcal{O}(1)$; for $\alpha>1$ the band width exceeds the gap, causing spectral overlap and interband mixing (see Supplementary Information).  Since higher bands broaden by an additional factor $\sim b$, the overlap criterion for band 1 already determines the onset of overlap for the full spectrum.

This band structure suggests a mapping in terms of hard-core bosons hopping on a lattice of size \( N \), where a boson  corresponds to spin up in the \( \sigma^x \) basis. 
Explicitly, Eq.~\eqref{HAM} takes the equivalent form of a bosonic Hamiltonian (see Methods for derivations):
\begin{align}
\label{effham0}
& \hat{H}_{\mathrm{eff}} = 
J_z \sum_i \left( \hat{a}_i^\dagger \hat{a}_{i-1} + \hat{a}_{i-1}^\dagger \hat{a}_i \right)
+ J_z \sum_i \left( \hat{a}_i^\dagger \hat{a}_{i-1}^\dagger + \hat{a}_i \hat{a}_{i-1} \right) \nonumber \\
& + \sum_i V(i) \,\hat{n}_i
+ \sum_{i<j} W_{i,j} \,\hat{n}_i \hat{n}_j
+ U \sum_i \hat{n}_i(\hat{n}_i-1)
+ E_{b=0} \,,
\end{align}
where
\begin{align*}
&V(i) = -2 J_x \sum_{j \ne i} \frac{1}{|i-j|^\alpha}, \;\;
&W_{i,j} = \frac{4 J_x}{|i-j|^\alpha}, \\
&U \rightarrow \infty \, \text{(hard-core)},\;\;
&E_{b=0} = J_x \sum_{i<j} \frac{1}{|i-j|^\alpha} \,.
\end{align*}
Here $V(i)$ acts as a local potential and $W_{i,j}$ represents the instantaneous density-density interaction between distant bosons. In this representation, the band number $b$ becomes the total boson occupation number $\sum_i \hat{n}_i$, and the second term in $\hat{H}_{\mathrm{eff}}$ contains the coupling between bands.

Manifestly, the only source of nonlocality in $\hat{H}_{\mathrm{eff}}$ is the two-body interaction 
$W_{i,j}\propto J_x/|i-j|^\alpha$. 
Indeed, if the $W$ term is removed from the Hamiltonian, the dynamics exhibit only local information spread (see Supplementary Information). \\

\textit{\textbf{Interband and intraband nonlocality and the emergence of nonlocal light cones---}}
\label{sec:inter-intra}
The long-range interaction term ($W$) drives two distinct nonlocal mechanisms: a programmable \textit{intraband} process operating strictly within a fixed excitation manifold, and an unprogrammable \textit{interband} process driven by virtual transitions between different manifolds.

The strengths of both mechanisms can be estimated by noting that the normalized Frobenius norm is exactly equal to the absolute difference in local boson densities: $|\langle \hat{n}_n^{(1)}(t) \rangle - \langle \hat{n}_n^{(2)}(t) \rangle|$ for two different initial states with a fixed boson number, see Supplementary Information. The {\it interband} nonlocality is generated by the second term in the r.h.s. of Eq.~{\eqref{effham0}}, which creates or annihilates a pair of bosons, giving rise to a transition from band $b$ to band $b \pm 2$. The probability to create or annihilate a boson pair will depend, through the $W$ term, on the distribution of bosons in the initial band-$b$ state.  In the presence of large band gaps, $\Delta E \sim J_x N^{1-\alpha} \gg J_z$, the interband signal can be modeled  via first-order time-dependent perturbation theory, yielding an interband signal strength that scales as (see Supplementary Information):
\begin{equation}
\label{Interband_main}
\|\Delta \rho_n(t)\|_F^{\mathrm{interband}}\sim \frac{\alpha J_z^2}{J_x^2 N^{3(1-\alpha)} r^{\alpha+1}}\,,
\end{equation}
where $r$ is the distance from the source.
Eq.~\eqref{Interband_main}, which is valid for $\alpha<1$, establishes that the interband background vanishes entirely in the thermodynamic limit ($N \to \infty$) or for strong long-range coupling ($J_x \to \infty$). The case  $\alpha=0$ is special and is discussed in detail in the Supplementary Information. 

In the limit of small interband nonlocality, information spread can be described by the band-projected effective Hamiltonian $H_{\mathrm{eff}}^b$, obtained from Eq.~(\ref{effham0}) by omitting the second term. In the projected Hamiltonian, which preserves the total boson number, the interplay of the kinetic hopping term proportional to $J_z$ and the density-density interaction $W_{i,j}$ determines the emergence of nonlocal light cones at the boson positions. Physically, when a local perturbation shifts the source boson, it instantly modulates the interaction potential felt by all remote background bosons. This sudden quench of the local energy landscape triggers a response: the remote bosons begin to propagate via the nearest-neighbor hopping term $J_z$, generating independent, localized light cones that erupt directly from their respective positions. 
This {\it intraband} signal can be estimated by computing
the time-evolved density operator 
$\hat n_{q}(t) = U^\dagger(t)\hat n_{q}U(t), \qquad U(t)=e^{-iH_{\mathrm{eff}}^bt}$. Expanding the evolution operator to fourth order in time, one obtains: 
\begin{align}
\label{Intraband_main}
\|\Delta \rho_n(t)\|_F^{\mathrm{intraband}}
=
\frac{t^4J_z^2(4J_x)^2}{3}
\alpha^2(\alpha+1)
\frac{1}{r_0^{2\alpha+3}} + \mathcal{O}(t^5) \,,
\end{align}
independent of both the system size $N$ and the specific band index $b$.

Fig.~\ref{NatureFig2} demonstrates how increasing the long-range interaction strength $J_x$ gives rise to a non-local light cone. At small $J_x$, information spreads unconstrained across the entire lattice [Fig.~\ref{NatureFig2}(a)], but as $J_x$ increases, a distinct non-local light cone originates at the target excitation [Fig.~\ref{NatureFig2}(b)]. This trend is confirmed by the fixed-time slice in Fig.~\ref{NatureFig2}(c): increasing $J_x$ suppresses the interband signal and amplifies the intraband signal, consistent with the scaling of Eqs.~(\ref{Interband_main}) and (\ref{Intraband_main}) shown in Fig.~\ref{NatureFig2}(d).

 
To characterize the signal at nonlocal light cones, we study ${\hat H}_{\mathrm{eff}}^{b}$.  In  Fig.~\ref{Prjb3}(a-c), the Frobenius norm is evaluated for two initial states in $b=3$ that differ   
by a displacement of the first excitation from site 2 to site 3, with the positions of the other two excitations held fixed. Fig.~\ref{Prjb3}(a-c) clearly shows how the causal space-time structure can be precisely shaped: nonlocal light cones erupt precisely at the positions of the fixed excitations and propagate 
independently, each exhibiting local behavior as seen in their respective profiles [Fig.~\ref{Prjb3}(d-f)]. 
In order to study how the intraband signal (inside nonlocal light cones) decays with distance, we consider a pair of initial states in band $b=2$ and vary the target excitation position. The intraband signal decays with distance with an exponent predicted by Eq.~(\ref{Intraband_main}), $p(\alpha) = 2\alpha + 3$, see Fig.~\ref{Prjb3}(g,h).  Also, the scaling with other parameters  predicted by Eqs.~(\ref{Interband_main},~\ref{Intraband_main}) is confirmed numerically: the $\alpha$-dependence for $\alpha<1$ is verified in  Fig.~\ref{N20LFIMb2VaryalphaJx5}(h) and the scaling with $J_x$ is verified in Fig.~\ref{NatureFig2}(d).  For the scaling with $N$ and more details, see Supplementary Information. 

Crucially, the emergence of nonlocal light cones requires a spatially dependent potential; for infinite-range interactions ($\alpha=0$), $W$ reduces in each band to an overall spectral shift, and no nonlocal light cones emerge, see Supplementary Information. Also note that the nearest-neighbor interaction $J_z$ is vital to this architecture; in the absence of this term, or upon substituting it with a local external field, nonlocal (and local) light cones disappear entirely, see Supplementary Information.

\begin{figure*}[htbp!]
    \centering
    \includegraphics[width=\linewidth]{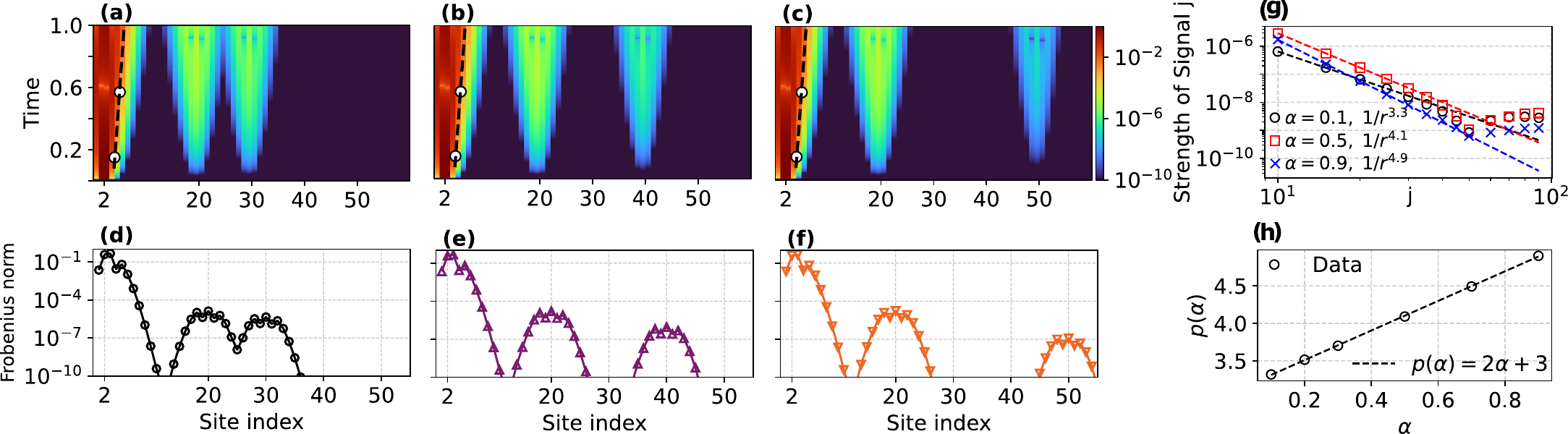}
    \caption{
    Density plots of the Frobenius norm showing light-cone dynamics under the projected Hamiltonian $\hat{H}_{\mathrm{eff}}^{b=3}$, obtained by varying the target position $j$ for initial states prepared in band $3$, $(2,20,j;\,3,20,j)$ (notation defined in Eq.~\eqref{pairnotation}), for $j=30$ \textbf{(a)}, $j=40$ \textbf{(b)}, and $j=50$ \textbf{(c)}. Other parameters are: $J_x=0.5$, $\alpha=0.5$, and $N=100$ (only $60$ sites are shown). 
    \textbf{(d,e,f)} The respective light-cone profiles at $t=1$. The amplitude of the nonlocal light cone at $j$ decreases as $j$ increases, while the nonlocal cone originating at site $20$ remains unaffected by the variation of $j$. 
    In panels \textbf{(g,h)}, $\hat{H}_{\mathrm{eff}}^{b=2}$ is used to study the decay of the intraband signal with $j$ using the initial states $(2,j;\,3,j)$ with $N=100$, $J_z=1$, and $J_x=0.1$. 
    \textbf{(g)} Power-law scaling of the intraband signal in the nonlocal light cone. Summing the signal over bulk sites (sites $10\text{--}90$) at $t=0.5$ and varying the excitation position $j$ yields a power-law scaling $\sim 1/r^{p(\alpha)}$, where $r = j - 3$ is the distance between message and target excitations. Note that when the distance exceeds $N/2$, the signal increases due to boundary conditions (see Supplementary Information). 
    \textbf{(h)} Extracted exponent $p$ as a function of $\alpha$, showing the relation $p(\alpha)=3+2\alpha$.
    }
    \label{Prjb3}
\end{figure*}

The competition between the intraband and interband signals determines the controllability of the causal landscape. By evaluating the intraband signal at the characteristic transport time $t \sim 1/J_z$, we identify an upper bound for the critical distance $r_c$ below which the causal landscape is controllable. Demanding that the intraband signal overcomes the static interband background, $F_{\mathrm{intraband}}(1/J_z) > F_{\mathrm{interband}}$, leads to the condition, for $\alpha<1$:
\begin{equation}
r_0^{\alpha+2} < \left(\frac{J_x}{J_z}\right)^4 N^{3(1-\alpha)}.
\end{equation}
This boundary defines the critical radius $r_c$. In the thermodynamic limit ($N \to \infty$), $r_c \to \infty$; the unprogrammable background noise is perfectly suppressed by the large energy gap $\Delta E \sim J_x N^{1-\alpha}$, making the entire chain controllable. In the Supplementary Information, a more detailed derivation is shown together with the critical radius in the presence of Kac rescaling.\\

\textit{\textbf{Emergence of locality---}}
\label{sec:Emergence}
Numerous studies have shown that the presence of long-range interactions does not guarantee enhanced information propagation; frequently, dynamics remain confined within a linear light cone for timescales that grow with system size~\cite{Cevolani_2016, Gong_2014, Kuwahara_2020, Luitz_2019, Feig_2015, arrufatvicente2024, Tan_2021, Mazza_2019, Kormos_2017, Defenu_2023, Defenu_2024, schachenmayer2013entanglement}. Some aspects of this phenomenon,  also termed \textit{cooperative shielding}, have been confirmed experimentally in trapped ion systems~\cite{Tan_2021}. Nevertheless a unified microscopic understanding of when and why locality emerges from inherently nonlocal Hamiltonians has remained elusive.

Our analysis clarifies that the emergence of local versus nonlocal behavior in long-range interacting systems is dictated directly by the initial conditions, a result that generalizes previous findings~\cite{Santos2016,Celardo_2016}. 

As derived above, the limit $N^{1-\alpha}J_x \gg J_z$ suppresses interband leakage. Dynamics are then dictated by the Hamiltonian projected onto manifolds of conserved excitation number, leaving the intraband density-density interaction $W_{i,j}$ as the unique source of nonlocality. Crucially, the role of $W_{i,j}$ depends on how these excitations are spatially distributed.

If the initial state consists of excitations spatially confined to a specific region, the nonlocal $W$ term vanishes outside this region. Consequently, information spread remains strictly local in the bulk, and nonlocal light cones emerge \textit{only} at the specific positions of the pre-existing excitations.
Conversely, if the initial condition contains excitations distributed across the entire chain (e.g., a global quench from a $z$-polarized state), the $W$ term connects all sites simultaneously, leading to highly nonlocal propagation, see Supplementary Information for a detailed comparison of states initialized in different bases.
This framework elegantly reconciles seemingly contradictory experimental results. \\

\textit{\textbf{Conclusions and discussion---}}
\label{sec:discussion}
We have demonstrated that nonlocality in long-range interacting systems can be precisely engineered, allowing the causal space-time landscape of many-body quantum spin systems to be programmably shaped.  By exploiting the interplay between long-range and short-range interactions, we established a protocol to induce nonlocal light cones at deterministic locations, effectively allowing information to bypass the geometric bulk of the system.

Our findings are immediately actionable in state-of-the-art quantum simulators. Trapped ion platforms are particularly well-suited to verify these predictions: while they natively realize long-range interactions, the specific Hamiltonian required here can be engineered using established digital or hybrid quantum simulation protocols~\cite{Monroe2021,lanyon2011universal,Korenblit2012}. Furthermore, recent advances in multimode cavity QED have enabled tunable-range and sign-changing interactions~\cite{Vaidya2018, Guo2019, Ritsch2013}, offering an alternative pathway to realize the connectivity required for our protocol.

Ultimately, engineered nonlocal light cones elevate connectivity to a programmable resource. By shaping where and how information propagates, our protocol enables targeted long-range communication without sacrificing local isolation. This opens new directions for quantum communication, state transfer, simulation, error correction, and distributed quantum computing, while providing a general framework for designing many-body systems with tailored information-propagation landscapes.\\

\section*{Methods}
\textbf{\textit{Conventions---}}
\label{Conventions}Throughout this work, we use a normalized Frobenius norm, defined by dividing
$\|\Delta \rho_n(t)\|_{F}$ by its global maximum over all sites and times, thus mapping all signals to the interval $[0,1]$. For brevity, we refer to this normalized quantity simply as the Frobenius norm.\\

\textit{\textbf{Numerical methods---}}\label{Numerics} Real-time dynamics are computed by direct integration of the time-dependent Schr\"odinger equation using a fourth-order Runge-Kutta (RK4) scheme. Time evolution is performed either in the full many-body Hilbert space or within a band-projected manifold, obtained by restricting the Hamiltonian to a suitably chosen manifold with a fixed excitation number.
The time step $\Delta t$ is chosen sufficiently small to ensure numerical convergence of all reported observables. Convergence is assessed by systematically reducing $\Delta t$ until relative changes in local observables, Frobenius norms, and extracted light-cone velocities between successive runs fall below numerical precision. Typical simulations employ $\Delta t \in [10^{-7},\,10^{-5}]$, depending on system size and interaction strength. While the RK4 scheme is not exactly unitary and accumulates errors at long times, all results reported in the main text correspond to evolution times $t \lesssim 10$, over which norm conservation is explicitly monitored at each time step, with the maximum deviation satisfying $|\langle\psi(t)\mid\psi(t)\rangle-1| < 10^{-14}.$

Exact diagonalization is employed to compute spectral quantities and to benchmark RK4 simulations. Initial states are chosen as product states with a localized perturbation in the $\sigma_x$ basis, prepared within a fixed excitation sector unless stated otherwise. All numerical results are cross-validated using independent diagnostics,  yielding consistent behavior across  different system sizes (further details can be found in the Supplementary Information).\\

\textit{\textbf{Formal derivation: Holstein–Primakoff transformation and linear spin wave theory---}} 
\label{HP_Derivation}
We derive the effective Hamiltonian for the spin model of Eq.~\eqref{HAM} using the Holstein–Primakoff (HP) transformation. Crucially, because spin-1/2 systems strictly forbid double occupancy, the higher-order corrections in the Holstein-Primakoff transformation identically vanish. Consequently, this mapping to a hard-core boson model is not restricted to the low-excitation regime, but is fully exact across all excitation manifolds.

The standard HP transformation assumes a fully $+z$ polarized ferromagnetic vacuum, and the HP mapping reads:
\begin{align}
    \hat{S}_z &= S - \hat{n}, \quad \\
\hat{S}_+ &= \sqrt{2S} \sqrt{1 - \frac{\hat{n}}{2S}}\, \hat{a}\,, \quad \\
\hat{S}_- &= \sqrt{2S} \hat{a}^\dagger \sqrt{1 - \frac{\hat{n}}{2S}} \,,
\end{align}
where $S$ denotes the total spin quantum number of the local magnetic moments, $\hat{n} = \hat{a}^\dagger \hat{a}$ is the local boson number operator, and the ladder operators are defined as  $\hat{S}_\pm = \hat{S}_x \pm i \hat{S}_y = \frac{1}{2}(\sigma_x \pm i \sigma_y)$.

In the low-excitation (dilute) regime, where $\hat{n} \ll 2S$, the square root can be expanded as a power series:
\begin{equation}
\label{sqrtExpansion}
    \sqrt{1 - \frac{\hat{n}}{2S}} = 1 - \frac{\hat{n}}{4S} - \frac{\hat{n}^2}{32S^2} - \cdots.
\end{equation}
In the zeroth-order (linearized) approximation of the HP transformation, which keeps only the leading behaviour in $\langle \hat{a}^\dagger \hat{a} \rangle$, the spin operators become $\hat{S}_+^{(0)} = \sqrt{2S} \, \hat{a}, \quad
\hat{S}_-^{(0)} = \sqrt{2S} \, \hat{a}^\dagger, \quad
\hat{S}_z^{(0)} = S - \hat{a}^\dagger \hat{a}$, and so $\sigma_x \approx \hat{a} + \hat{a}^\dagger, \quad
\sigma_y \approx i(\hat{a}^\dagger - \hat{a}), \quad
\sigma_z = 1 - 2 \hat{a}^\dagger \hat{a}.$

 However, our model is defined with a ground state polarized along the $x$-axis, so we apply a $\pi/2$ rotation about the $y$-axis, $\hat{R}_y\left( \tfrac{\pi}{2} \right) = e^{-i \frac{\pi}{4} \sigma_y}.$ Substituting the linearized HP expressions into the rotated frame gives $\sigma_z \approx -(\hat{a} + \hat{a}^\dagger), \quad
\sigma_y \approx i(\hat{a}^\dagger - \hat{a}), \quad
\sigma_x = 1 - 2\hat{n}.$ Note that the terms involving \(\sigma_x\) (i.e., the \(J_x\) terms) are exact. The hopping terms, on the other hand, are the zeroth-order contributions in the expansion shown in Eq.~\eqref{sqrtExpansion}.

Applying the rotated bosonic operators to the Hamiltonian of Eq.~\eqref{HAM},
we obtain:
\begin{align}
\label{0thOrder}
{\hat H}^{(0)} ={}& 
J_z \sum_j \left( \hat{a}_j^\dagger \hat{a}_{j+1} + \hat{a}_{j+1}^\dagger \hat{a}_j + \hat{a}_j^\dagger \hat{a}_{j+1}^\dagger + \hat{a}_j \hat{a}_{j+1} \right) \nonumber \\
& -2 \sum_{i<j} \frac{J_x}{|i-j|^\alpha} (\hat{n}_i + \hat{n}_j) 
+ 4 \sum_{i<j} \frac{J_x}{|i-j|^\alpha} \hat{n}_i \hat{n}_j \nonumber \\
& + \sum_{i<j} \frac{J_x}{|i-j|^\alpha}.
\end{align}

Beyond linear order, the HP expansion introduces density-dependent hopping terms $\sqrt{1 - \frac{\hat{n}}{2S}} \approx 1 - \frac{\hat{n}}{4S}$, leading to interaction-assisted hopping, which we denote \( \delta \hat{H}^{(1)} \). 
So 
\begin{equation}
    {\hat H}= {\hat H}^{(0)} +  \delta \hat{H}^{(1)} \,,
\end{equation}
where ${\hat H}^{(0)}$ is given by Eq.~\eqref{0thOrder}, and the additional terms at the next order are:
\begin{align*}
\delta \hat{H}^{(1)} = J_z \sum_{j=1}^{L-1} \Bigg[-\frac{1}{4} (\hat{a}_j + \hat{a}_j^\dagger)(\hat{n}_{j+1} \hat{a}_{j+1} + \hat{a}_{j+1}^\dagger \hat{n}_{j+1}) \\- \frac{1}{4} (\hat{n}_j \hat{a}_j + \hat{a}_j^\dagger \hat{n}_j)(\hat{a}_{j+1} + \hat{a}_{j+1}^\dagger) \\+ \frac{1}{16} (\hat{n}_j \hat{a}_j + \hat{a}_j^\dagger \hat{n}_j)(\hat{n}_{j+1} \hat{a}_{j+1} + \hat{a}_{j+1}^\dagger \hat{n}_{j+1})
\Bigg].
\end{align*}

Since in spin-$\tfrac{1}{2}$ ($S = 1/2$) systems double occupancy is forbidden, every term in $\delta \hat{H}^{(1)}$ necessarily attempts either over-annihilation or creation on an already occupied site, and therefore all such contributions vanish identically in every band. Thus we recover the effective Hamiltonian, Eq.~\eqref{effham0}.\\

\textit{\textbf{Subspace projection derivation of the effective Hamiltonian---}}\label{projection_derivation} We begin with the single–excitation band ($b=1$). The nearest–neighbor Ising term induces hopping, giving

\begin{equation}
\label{tight_binding2}
\hat {H}_{\mathrm{eff}}^{b=1} = J_z \sum_j \left(|j\rangle\langle j+1| + \mathrm{h.c.}\right)
- 2 J_{x} \sum_j \epsilon_j |j\rangle\langle j|.
\end{equation}

Thus the $b=1$ band maps to a local tight–binding model with position-dependent onsite potential.  \\


\noindent\textbf{Higher bands---}  
A basis state in band $b$ has $b$ flipped spins. Let us define the set of positions $D$ with  spins flipped, e.g., $D=\{3,4,7\}$ for $b=3$ if spins in position ${3,4,7}$ are flipped. The contribution to the energy from the $J_{x}$ interaction can be decomposed as 
\begin{widetext}
\begin{eqnarray}
E &=& J_{x} \sum_{i<j} \frac{ \sigma_i^x  \sigma_j^x}{(j-i)^{\alpha}} = J_{x} \left[\sum_{\substack{i<j \\ i,j \not\in D}} \frac{1}{(j-i)^{\alpha}}
-\sum_{\substack{i<j \\i \in D,j \not\in D}} \frac{1}{(j-i)^{\alpha}}
-\sum_{\substack{i<j \\i \not\in D,j \in D}} \frac{1}{(j-i)^{\alpha}}
+\sum_{\substack{i<j \\i,j\in D}} \frac{1}{(j-i)^{\alpha}} \right] \nonumber \\
&=& J_{x} \left[ \sum_{i<j} \frac{1}{(j-i)^{\alpha}} 
-2 \sum_{j \in D} \sum_{i \ne j} \frac{1}{|j-i|^{\alpha}} 
+4 \sum_{\substack{i,j \in D \\ i<j}} \frac{1}{|j-i|^{\alpha}} \right] = E_{b=0} + \sum_{j \in D} V(j) + \sum_{\substack{i,j \in D \\ i<j}} W_{i,j},
\end{eqnarray}
\end{widetext}
where
$$E_{b=0}= J_{x} \sum_{i<j}\frac{1}{(j-i)^{\alpha}},\quad
V(j)=-2J_{x} \sum_{i\neq j}\frac{1}{|j-i|^{\alpha}},$$
and
$$W_{i,j}=\frac{4J_{x}}{|j-i|^{\alpha}}.$$

Thus, the projected Hamiltonian for band $b$ contains a local onsite potential $V$ and a nonlocal two-body term $W$.  

It is interesting to note that in the case of  Kac rescaling (explicitly addressed in the Supplementary Information), for which $J_x \sim J_{\mathrm{long}}/N^{1-\alpha}$, the one-body potential scales as $O(\alpha J_{\mathrm{long}})$ for small $\alpha$, while the two-body term is at most $O(b^2 J_{\mathrm{long}}/N^{1-\alpha})$. Hence for $\alpha<1$, fixed $b$, and $N\to\infty$, $W$ becomes negligible and the effective Hamiltonian approaches locality.  \\

\noindent\textbf{Second–quantized form---}  
Introducing creation and annihilation operators $\hat a_i^\dagger,\hat a_i$ for up spins in the $\sigma_x$ basis, one finds
\begin{equation}
\hat V = \sum_i V(i)\hat a_i^\dagger\hat a_i 
+ \sum_{i<j} W_{i,j}\hat a_i^\dagger \hat a_i \hat a_j^\dagger \hat a_j ,
\end{equation}
with a hard–core constraint forbidding double occupancy. The Ising term provides the kinetic energy
\begin{equation}
\hat T = J_z \sum_i \left(\hat a_i^\dagger \hat a_{i-1} + \hat a_{i-1}^\dagger \hat a_i\right).
\end{equation}
Collecting terms gives the effective Hamiltonian \eqref{effham0}, a hard-core Bose–Hubbard model with one-body and two-body potentials. All contributions are local except for $W$.\\

\medskip
\textbf{Data and code availability:} Code and data supporting the findings of this study are available from the authors upon reasonable request.\\

\textbf{Author contributions:} S.S. performed all numerical simulations and carried out most of the analytical derivations. G.L.C. and F.B.  supervised the project and contributed to the analytical understanding of nonlocal signal propagation. L.K. supervised the research and introduced the boson mapping and the Frobenius norm framework, which enabled the interpretation of nonlocal signal dynamics. All authors contributed to writing and editing the manuscript.\\

\textbf{Acknowledgments:} This research was supported in part using high-performance computing (HPC) resources and services provided by Information Technology at Tulane University, New Orleans, LA.\\

\textbf{Competing interests:} The authors declare no competing interests.\\

\textbf{Additional information:} Supplementary Information is available for this paper.

\bibliographystyle{apsrev4-2}
\bibliography{references}

\clearpage
\pagebreak

\onecolumngrid
\section*{Supplementary Information: \\Shaping causality: programmable nonlocal signal generation in long-range spin systems}
\appendix 
\twocolumngrid
\renewcommand{\thesection}{S\arabic{section}}
\renewcommand{\theequation}{S\arabic{equation}}
\renewcommand{\thefigure}{S\arabic{figure}}
\renewcommand{\thetable}{S\arabic{table}}

\setcounter{page}{1}
\setcounter{section}{0}
\setcounter{equation}{0}
\setcounter{figure}{0}

\renewcommand{\addcontentsline}[3]{\oldaddcontentsline{#1}{#2}{#3}}
\tableofcontents

\section{Phenomenology and Numerics}
\subsection{Other Information Spread Figures of Merit}
\label{OtherMeasures}

\begin{figure}[htbp!]
    \includegraphics[width=\linewidth]{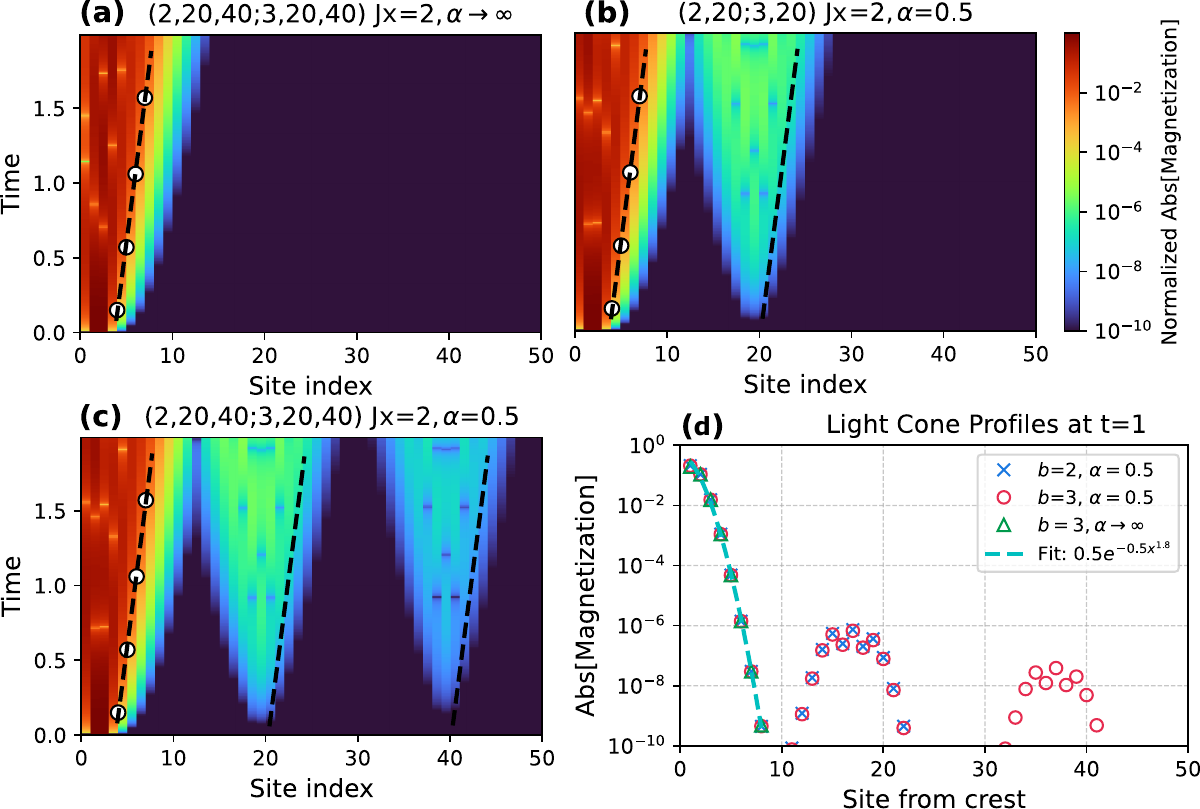}
    \caption{\textbf{(a,c)} Light-cone dynamics under the projected Hamiltonian ${\hat H}_{\mathrm{eff}}^{b=3}$  for band-3 initial states $(2,20,40;\,3,20,40)$ (notation defined in Eq.~\eqref{pairnotation} in the main text), and \textbf{(b)} ${\hat H}_{\mathrm{eff}}^{b=2}$  for the band-2 state $(2,20;\,3,20)$, to which the full Hamiltonian converges in the large-$J_x$ limit. Dynamics are computed from the site-resolved magnetization
    $\left|\frac{\bra{\psi_{1}}\sigma^x_n\ket{\psi_{1}}-\bra{\psi_{2}}\sigma^x_n\ket{\psi_{2}}}{2}\right|$,
    shown for $\alpha\!\to\!\infty$ \textbf{(a)} and $\alpha=0.5$ \textbf{(b,c)} ($\alpha\!\to\!\infty$ implemented by retaining only nearest-neighbor couplings). Parameters: $J_x=2$, $N=100$ (first 50 sites shown). \textbf{(d)} Light-cone profiles at $t=1$ extracted from (a–c). These observables reproduce the Frobenius-norm results. In \textbf{(a)}, no secondary light cones appear due to purely nearest-neighbor couplings. In \textbf{(c)}, the central cone evolves independently of the right cone, consistent with expectations for non-interacting bosons. Similar behavior appears in other observables, e.g., correlation functions (Fig.~\ref{Correlation}).}
    \label{NvariedLFIMHeffb3lcalpha05farMagPlot}
\end{figure}

\begin{figure}[htbp!]
    \centering
    \includegraphics[width=1\linewidth]{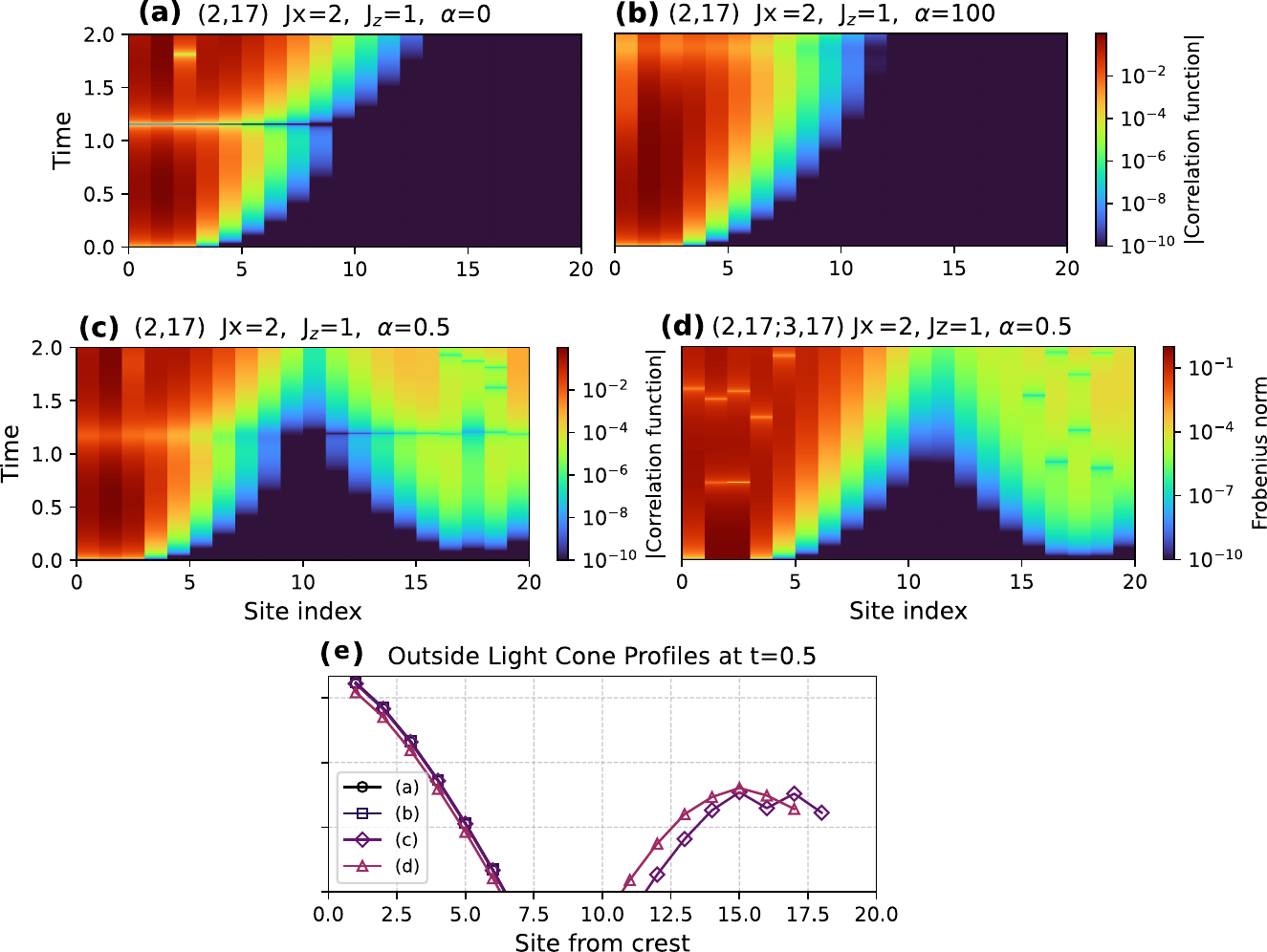}
    \caption{\textbf{(a–c)} Light-cone dynamics under the projected Hamiltonian $\hat H_{\mathrm{eff}}^{b=2}$, computed using the connected two-point correlator $|\langle \sigma_2^x(t)\sigma_j^x(t)\rangle
    - \langle \sigma_2^x(t)\rangle \langle \sigma_j^x(t)\rangle|$
    for the band-2 initial state $(2,17)$ (notation in Eq.~\eqref{pairnotation} in the main text). Panels (a-c) show the results for three different values of the exponent $\alpha$, for $N=20$ and $J_x=2$. Note that the apparent signal dips at $t\approx 1.2$ are artifacts of taking the absolute value, occurring where the correlation function changes sign. \textbf{(d)} Frobenius-norm dynamics with $\alpha=0.5$ for the state $(2,17;\,3,17)$, demonstrating that the Frobenius-norm signatures can be reproduced using experimentally accessible two-point correlations. \textbf{(e)} Light-cone profiles at $t=0.5$ extracted from panels \textbf{(a–d).}}. 
    \label{Correlation}
\end{figure}

In the main text, we investigated information spread using the Frobenius norm  which, although not directly experimentally measurable, provides the most robust and widely used characterization of information spread. Unlike two-point correlators, out-of-time-ordered correlators, or magnetization, which probe specific observables in fixed bases, the Frobenius norm in Eq.~\eqref{eq:fnorm} in the main text, captures the maximal change in measurement outcomes and is therefore sensitive to coherent differences that other probes may miss. To connect this with  more experimentally accessible   quantities, let us  consider the site-resolved magnetization difference (Fig.~\ref{NvariedLFIMHeffb3lcalpha05farMagPlot}),
\begin{equation}
M_n(t)=\left|\frac{\langle\psi_1|\sigma_n^x(t)|\psi_1\rangle-
\langle\psi_2|\sigma_n^x(t)|\psi_2\rangle}{2}\right|,
\label{eqsi_mx}
\end{equation}
and the connected two-point correlator (Fig.~\ref{Correlation}),
\begin{equation}
C_{2j}(t)=\left|\langle\sigma_2^x(t)\sigma_j^x(t)\rangle-
\langle\sigma_2^x(t)\rangle\langle\sigma_j^x(t)\rangle\right|.
\label{eqsi:cor}
\end{equation}
In Eq.~(\ref{eqsi:cor}), the average $\langle...\rangle$ is taken over the evolved state $\ket{\psi_1(t)}$, so that in this approach only one initial state is needed, in contrast with Eq.~(\ref{eqsi_mx}).
 Figure~\ref{NvariedLFIMHeffb3lcalpha05farMagPlot} shows the site-resolved magnetization difference. Figure~\ref{Correlation}  considers instead a single band-2 state $(2,17)$ and tracks correlation spread from site~2. Despite the different observables, the propagation fronts display the same qualitative behavior as the Frobenius norm shown in the main text. These results show that the tunable propagation identified via the Frobenius norm, controlled by $J_x$, excitation separation ($r=i-j$), and $\alpha$, can also be observed with experimentally accessible observables. Such tunability is absent in short-range Hamiltonians.

\newpage
\subsection{Light-Cone Velocity}
\label{LCvelocity}

In this section, we analytically show the light-cone velocity of our model to be  $2J_z$ for the Hamiltonian projected on a fixed-$b$ manifold. The projected Hamiltonian $H_{\mathrm{eff}}^b$ can be obtained from the effective Hamiltonian, Eq.~\eqref{effham0} in the main text, with the interband coupling turned off,
\begin{align}
\hat H_{\mathrm{eff}}^b 
&= J_z \sum_i \!\left( \hat a_i^\dagger \hat a_{i-1} + \hat a_{i-1}^\dagger \hat a_i \right)
+ \sum_i V(i)\, \hat n_i \nonumber\\
&\quad + \sum_{i<j} W_{i,j}\, \hat n_i \hat n_j
+ U \sum_i \hat n_i(\hat n_i+1)   \,,
\end{align}
where $b=\sum_i n_i$.\\

Moreover, we impose periodic boundary conditions so that 
\(V(i)=V,\; W_{i,j}=W(|i-j|),\; U=\text{const}\).

For $b=1$, transport is only governed by the nearest-neighbor hopping term:
\begin{equation}
\hat H_{\text{hop}} = J_z \sum_i \!\left( \hat a_i^\dagger \hat a_{i-1} + \hat a_{i-1}^\dagger \hat a_i \right).
\end{equation}

Using the Fourier transform 
\[
\hat a_j = \frac{1}{\sqrt{N}}\sum_k e^{\mathrm{i}kj}\hat a_k, \qquad 
\hat a_j^\dagger = \frac{1}{\sqrt{N}}\sum_k e^{-\mathrm{i}kj}\hat a_k^\dagger,
\]
and orthogonality 
\(\frac{1}{N}\sum_j e^{\mathrm{i}(k-k')j}=\delta_{k,k'}\),
we obtain a diagonal form in momentum space
\[
\hat H_{\text{hop}} = 2J_z \sum_k \cos k\, \hat a_k^\dagger \hat a_k.
\]

The corresponding group velocity follows directly from the dispersion relation,
\begin{equation*}
v_g(k)=\frac{dE(k)}{dk}=-2J_z\sin k,
\end{equation*}
and is maximized at $k=\pm\pi/2$, giving
\begin{equation}
v_g^{\max}=2|J_z|.
\end{equation}
This velocity is consistent with that observed in Fig. \ref{N20LFIMb2VaryalphaJx5} in the main text.
The onsite terms $V$, $U$ merely shift the spectrum and thus do not affect \(v_g\).

In the large-$\alpha$ limit, the Hamiltonian \eqref{HAM} in the main text maps exactly to the XY Hamiltonian, whose dispersion relation is \cite{Farreras_2025, Butterfly_Velocity_XY_2024}:
\begin{equation*}
\epsilon(k) = -2\sqrt{J_x^2 + J_z^2 + 2J_x J_z \cos(2k)},
\end{equation*}
and the corresponding group velocity is
\begin{equation*}
v_g(k) = \frac{4 J_x J_z \sin(2k)}{\sqrt{J_x^2 + J_z^2 + 2J_x J_z \cos(2k)}} .
\end{equation*}
Maximizing $v_g(k)$  yields the Lieb–Robinson velocity 
\begin{equation}
v_{LR}= 4\,\min(J_x, J_z).
\end{equation}

This velocity is in very good agreement with the results shown in Figs. \ref{N20LFIMb2VaryalphaJx5}, \ref{NatureFig2}, and \ref{Prjb3} in the main text.

\subsection{Decay Outside the Light Cone}
\label{Outside-light-cone decay}

Light-cone velocities are extracted from first-arrival times, defined as the earliest times at which $\|\Delta \rho_n(t)\|_F \ge 0.01$, and obtained by linear fits to these arrival fronts. Outside-the-cone profiles are evaluated at fixed times by locating the outward-propagating wave crest and fitting the decay of $\|\Delta \rho_n(t)\|_F$ beyond it (see insets in Fig.~\ref{MidBandInitN15b7Wneq0alongzPlot}). While power-law fits in the exponent provide a convenient description, previous work on related tight-binding models has shown that the asymptotic decay is expected to follow a form of the type $\exp[-x \log x]$ \cite{Chen2023SpeedLimits}. 

To compare these possibilities, we fit the logarithm of the signal using two functional forms:
\begin{align}
\log y &= -k x^m, \\
\log y &= -x\log(ax) + b .
\end{align}

\begin{figure}[htbp!]
\centering
\includegraphics[width=\linewidth]{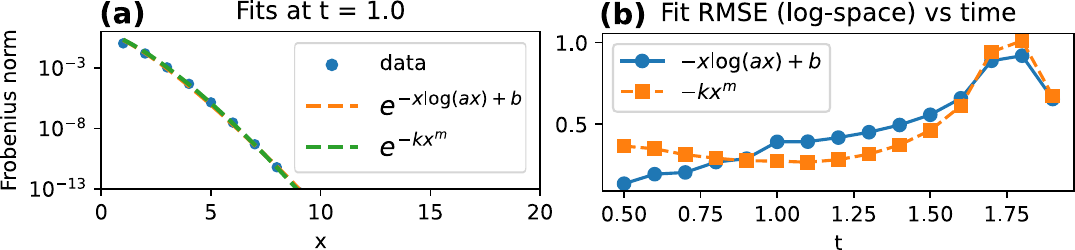}
\caption{
Comparison of functional forms describing the decay of the signal outside the light cone. Here $x$ denotes the distance from the wave crest.
\textbf{(a)} 
Example profile at $t=1$ under the projected Hamiltonian $\hat H_{\mathrm{eff}}^{b=1}$ with $J_x=2$, $\alpha=0$, and $N=20$ for an initial state with an excitation at site~2. The data are fitted using $\exp[-x\log(ax)+b]$ and $\exp(-k x^m)$ with a=2.9, b=-0.5, k=1.5, m=1.4.
\textbf{(b)} Root-mean-square error (RMSE) of the two fits in log space as a function of time. The comparable RMSE values indicate that both models capture the qualitative decay behavior of the signal outside the light cone. This analysis is not restricted to band 1; since the bosons in our model are non-interacting, similar results hold for any number of excitations.}
\label{decay}
\end{figure}

In Figure~\ref{decay}, we compare the two different fits. Both models capture the qualitative decay of the profile outside the light cone. We therefore evaluate their relative performance by tracking the fitted parameters as a function of time and by comparing the root-mean-square error (RMSE) of the fits in log space. In practice, the RMSE values are comparable for the two models across the time window considered, indicating that both provide similarly good descriptions of the numerical data. Since the form $\exp(-k x^m)$ is simpler to parameterize and interpret, we use it as the default fitting form in the remainder of the analysis.

\subsection{Band Gap and Band Width}
\label{bandwidthgap_ratio}
For $\alpha=0$, the long-range interaction reduces to an all-to-all coupling. In this case, the Hamiltonian can be expressed in terms of the collective spin $S^x$, and the band energies  and  energy gaps  are given by
\begin{align}
    E_b &= \frac{J_x}{2}(N - 2b)^2 - \frac{J_x N}{2}, \\
    \Delta E_b &= E_b - E_{b-1} = -2 J_x \left(N-2b + 1\right).
\end{align}
Thus, the spectrum is quadratic in $b$, with band gaps linear in $b$. The largest gap is between $b=0$ and $b=1$ (or equivalently between $b=N$ and $b=N-1$), scaling as $\Delta E \sim 2J_x N$ for large $N$.\\

For nonzero $\alpha$, the 
largest gap scales as $\Delta E\sim 2J_x N^{\,1-\alpha}.$ Additionally, the intraband degeneracy is lifted. Let us consider  a basis state with the $k$-th spin flipped: $|k\rangle = \ket{+ \cdots \underbrace{-}_k \cdots +}$ in  the $b=1$ band. Each such state is an eigenstate of the long–range term, with energy
\begin{equation}
    E_k = E_{b=0} - 2 J_x \sum_{j\neq k}\frac{1}{|j-k|^{\alpha}}
         = E_{0} - 2 J_x\,\epsilon_k ,
\end{equation}
where the sum over all pairs involving site $k$ is:
\begin{equation}
\label{epsilonk}
    \epsilon_k = \sum_{s=1}^{k-1}\frac{1}{s^{\alpha}} + \sum_{s=1}^{N-k}\frac{1}{s^{\alpha}}.
\end{equation}
This defines a smooth external potential $\epsilon_k$. For $\alpha>1$, the sum converges to the Riemann zeta function, $\epsilon_k \to \zeta(\alpha)$.  
For $\alpha<1$, the sum diverges for large $N$ as $\epsilon_k \;\sim\; \frac{N^{\,1-\alpha}}{1-\alpha}\, $. The largest intraband energy difference (band width) occurs between a spin flip at the chain center and one at the edge, and is given by \(2 J_x\,\delta_{\alpha,N}\), where

\begin{equation}
\label{bandgap1}
    \delta_{\alpha,N} = \epsilon_{N/2} - \epsilon_{1} \approx \frac{N^{\,1-\alpha}}{1-\alpha}\,(2^{\alpha}-1) \,,
\end{equation}

which for large $N$ and small $\alpha$ behaves as
\begin{equation}
\delta_{\alpha,N} \;\approx\; \alpha(\ln 2)\, N^{\,1-\alpha}.
\end{equation}

Thus, the band width for the $b=1$ band
is
\begin{equation}
\delta E_1 \;\sim\; 2 J_x\, \delta_{\alpha,N}
\;\sim\; 2 J_x \alpha(\ln 2) N^{\,1-\alpha}.
\end{equation}

Comparing this with the inter-band gap, the ratio for small $\alpha$ is
\begin{equation}
\frac{\delta E_{1} }{\Delta E_{1}} 
\;\sim\; \frac{2 J_ x(\ln 2)\alpha N^{\,1-\alpha}}
{2J_x N^{\,1-\alpha}}
\;=\; (\ln 2)\alpha.
\label{bandwidth_bandgap}
\end{equation}
As $\alpha \to 1$, the ratio becomes of order unity. 

\begin{figure}[htbp!]
    \centering
    \includegraphics[width=\linewidth]{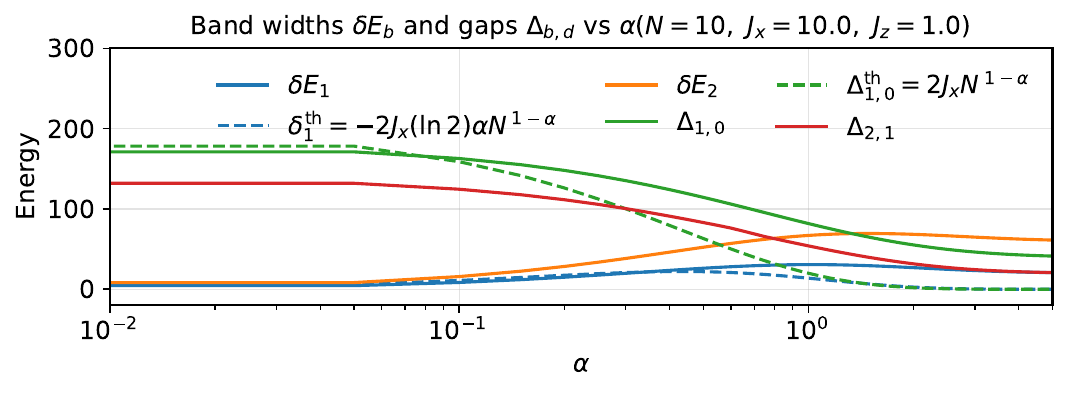}
    \includegraphics[width=\linewidth]{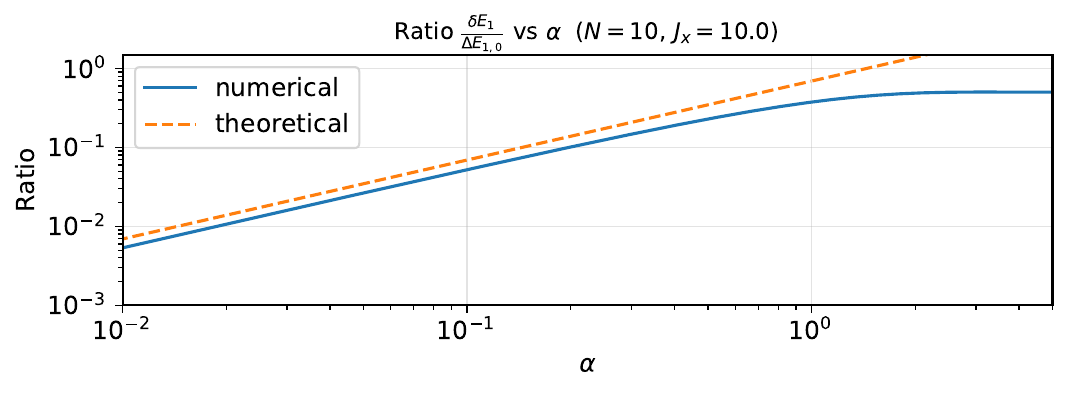}
    \caption{
    \textbf{Top:} Band widths $\delta E_b$, computed as the difference between the largest and smallest eigenenergies within band $b$, and band gaps $\Delta_{b,d}$, computed as the difference between the mean eigenenergies of bands $b$ and $d$, plotted as a function of $\alpha$ for $J_x=10$, $N=10$. Solid lines show numerical results, while dashed lines are the small-$\alpha$ theoretical predictions. 
    \textbf{Bottom:} Ratio of band width to band gap, $\delta E_b / \Delta_{b,d}$ becomes O(1) at $\alpha\sim1$, confirming the scaling predicted in Eq.~\eqref{bandwidth_bandgap}.
    }
  \label{band_metrics2}
\end{figure}

This behavior is verified numerically in Fig.~\ref{band_metrics2}. The top panel illustrates the transition of both the band widths $\delta E_b$ and the corresponding band gaps between band $b$ and $d$, $\Delta_{b,d}$  as a function of the long-range exponent $\alpha$, showing excellent agreement with our small-$\alpha$ analytical scaling (dashed lines). 

For higher bands, the band width is further enhanced by an extra factor of $b$, with all $b$ excitations located near the chain center and all $b$ excitations near the edge being the two extreme cases. 

Consequently, analyzing the onset of overlap for band 1 is sufficient, as it implies that all higher bands have already entered the overlapping regime. As a consistency check, the $b=2$ case is also included in Fig.~\ref{band_metrics2} (top); notice that the band width $\delta E_2$ merges with the gap $\Delta_{2,1}$ at a smaller $\alpha$ than the corresponding overlap for band 1. Thus, we can track the breakdown of the isolated band structure simply by plotting the ratio $\delta E_1 / \Delta_{1,0}$ as it approaches $\mathcal{O}(1)$ near $\alpha \sim 1$ in Fig.~\ref{band_metrics2} (bottom).



\subsection{Confinement Potential}
\label{confinement}

\begin{figure}[h]
    \centering
    \includegraphics[width=\linewidth]{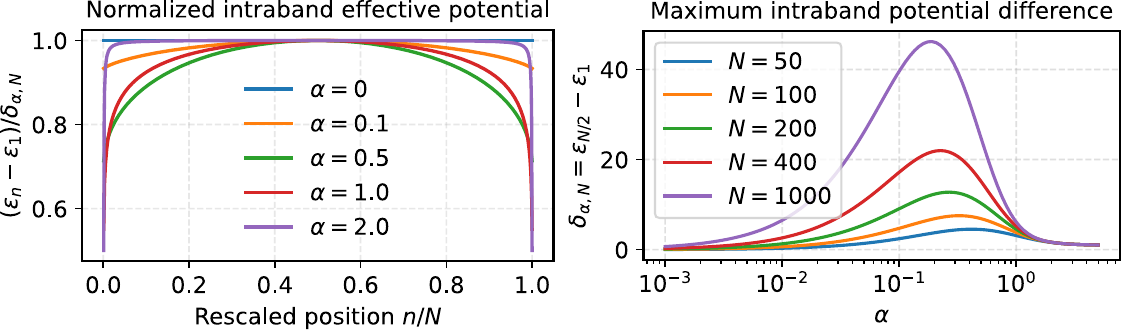}
    \caption{
    \textbf{Left:} Normalized intraband effective potential 
    $(\epsilon_n - \epsilon_1)/\delta_{\alpha,N}$
    as a function of the rescaled lattice position $n/N$ for $N=1000$, where $\delta_{\alpha,N} = \epsilon_{N/2} - \epsilon_{1}$ is the bandgap defined in Eq.~\eqref{bandgap1}. 
    For small $\alpha$, the potential is nearly flat, while larger $\alpha$ values produce a smooth confining profile 
    that pulls the single-spin excitation toward the chain center, 
    reflecting the emergence of localized intraband structure as long-range interactions strengthen.
    \textbf{Right:} Maximum intraband potential difference 
    $\delta_{\alpha,N}$ 
    as a function of the decay exponent $\alpha$. 
    For small $\alpha$, long-range interactions are nearly uniform and the potential is flat 
    ($\delta_{\alpha,N} \!\approx\! 0$). 
    As $\alpha$ increases,
    $\delta_{\alpha,N}$ grows, reaching a peak at some $\alpha<1$ and becoming $N$-independent in the short range regime $\alpha \gg 1$.     
    }
    \label{DeltaEpsilonVsAlpha}
\end{figure}

In the large-$J_x$ regime, excitation bands become well separated, making the band-projected Hamiltonian
[Eq.~\eqref{effham0} in the main text with the second term removed] quantitatively accurate to describe information propagation. Within a fixed band, the effective dynamics is governed by nearest-neighbor hopping with amplitude $J_z$ and a position-dependent intraband potential
\begin{equation}
V(i) = -2 J_x \sum_{j\neq i} \frac{1}{|i-j|^{\alpha}} .
\label{Vi}
\end{equation}
The hopping term $J_z$ alone sets the intrinsic propagation velocity. For $\alpha=0$, $V(i)$ is spatially uniform, the Hamiltonian is translationally invariant, and the light-cone velocity is analytically fixed to $v=2J_z$, independent of $J_x$, in agreement with both analytical and numerical results (Sec.~\ref{LCvelocity}).

For $\alpha>0$, long-range interactions lift the degeneracy of $V(i)$ and generate a spatially inhomogeneous, convex confining potential. This behavior is quantified in Fig.~\ref{DeltaEpsilonVsAlpha}, which shows the intraband potential profile and its dependence on the decay exponent $\alpha$ (left panel). For small $\alpha$, the potential remains nearly flat, while increasing $\alpha$ produces a smooth confining structure centered about the middle of the chain. The corresponding maximum intraband potential difference (right panel) grows with $\alpha$, peaks at some $\alpha<1$, and then decreases and becomes $N$-independent for large $\alpha$. 

\begin{figure}[htbp!]
    \centering
    \includegraphics[width=1\linewidth]{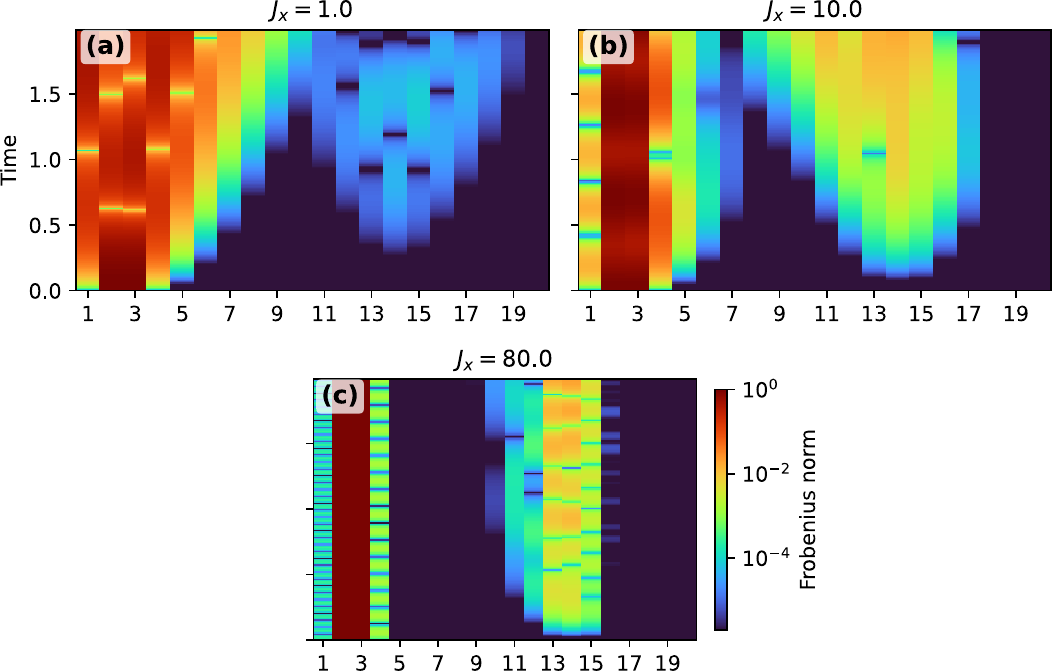}
    \caption{Density plots of the Frobenius norm showing light-cone dynamics generated by the band-projected Hamiltonian $\hat H_{\mathrm{eff}}^{b=2}$ for initial states prepared in band $b=2$, $(2,14;\,3,14)$ (notation defined in Eq.~\eqref{pairnotation} in the main text), for $N=19$ and $\alpha=0.5$. Increasing $J_x$ \textbf{(a-c)} deepens the effective confining potential $V(i)$, progressively spatially confining transport. The resulting light cone is asymmetric: propagation toward the center of the chain extends over more sites than propagation toward the edges, a direct consequence of the convex structure of $V(i)$.}
    \label{confinementLC}
\end{figure}

When $J_x \gg J_z$, the curvature of $V(i)$ increases linearly with $J_x$, and the resulting energy mismatch between neighboring sites can exceed the hopping scale $J_z$. In this regime, hopping is strongly suppressed, leading to real-space confinement. This explains why increasing $J_x$ enhances confinement for $\alpha>0$, while no such effect occurs for $\alpha=0$. This potential was also observed in \cite{Lerose2019quasilocalized, Monroe2021}.

The resulting confinement is directly visible in the light-cone dynamics of the band-projected Hamiltonian. Figure~\ref{confinementLC} shows Frobenius-norm density plots for dynamics projected onto band $b=2$ as $J_x$ is increased. As the confining potential deepens, transport becomes progressively localized. The light cone is also asymmetric: propagation toward the center of the chain extends over more sites than propagation toward the edges, a direct consequence of the convex, spatially inhomogeneous structure of $V(i)$.
 
\subsection{Emergence of Locality and Role of Nonlocal Term $W$}
\label{W0}
\begin{figure}[htbp!]
    \centering
    \includegraphics[width=\linewidth]{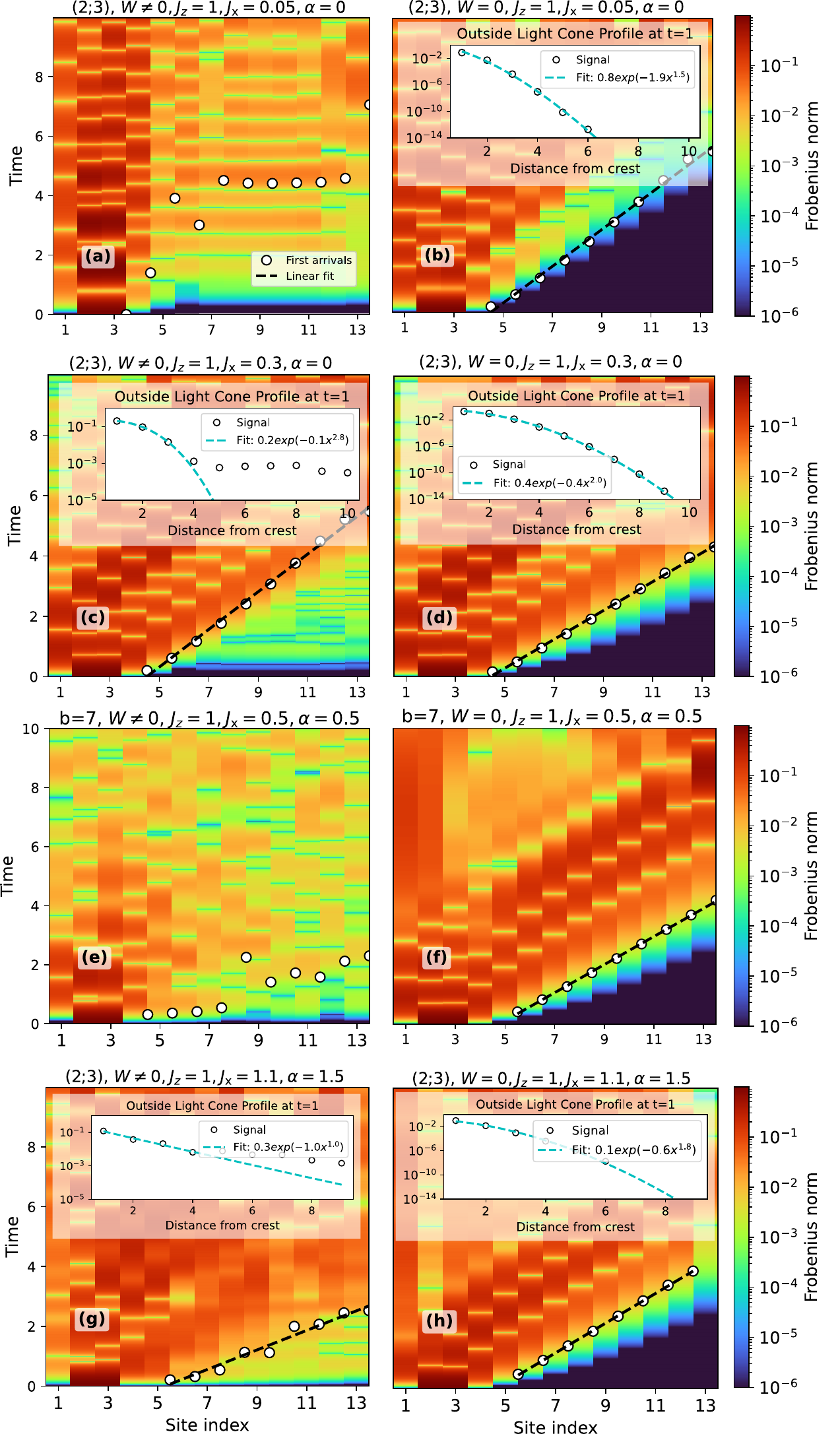}  
    \caption{Density plots of the Frobenius norm showing light-cone dynamics under the Hamiltonian \eqref{effham0} in the main text, for $N=13$ with different values of $J_x$ and $\alpha$.
    The left column corresponds to setting $W\ne 0$ (the full Hamiltonian dynamics), while in the right column, we set $W=0$.
    Setting $W=0$  enforces strict locality even for $N^{1-\alpha}J_x/J_z<1$, for initializations in high-lying bands, and for $1<\alpha<3$.  
    \textbf{(a,b)} Band~1 initialization at $N^{1-\alpha}J_x/J_z<1$: dynamics are nonlocal for the full Hamiltonian but strictly local when $W=0$. 
    \textbf{(c,d)}  Band~1 initialization for $N^{1-\alpha}J_x/J_z>1$: dynamics are quasi-local for the full Hamiltonian and strictly local with $W=0$. 
    \textbf{(e,f)} Initialization in a high-lying band, band~7, $(2,10,11,12,13,14,15;\,3,10,11,12,13,14,15)$, likewise shows nonlocal behaviour for the full Hamiltonian and locality when $W=0$. \textbf{(g,h)} Even when $\alpha=1.5$ with band~1 initialization, setting $W$ to 0 yields strictly local dynamics. Insets show the outside-the-cone profile at the time $t=1$.}   \label{MidBandInitN15b7Wneq0alongzPlot}
\end{figure}

In this section, we isolate the origin of nonlocal dynamics by comparing the evolution under the full Hamiltonian~\eqref{HAM} in the main text with one in which the nonlocal term $W$ is artificially removed. Specifically, we set $W=0$ in the Hamiltonian of Eq.~(\ref{effham0}) in the main text.

 Figure~\ref{MidBandInitN15b7Wneq0alongzPlot} compares the full dynamics ($W\neq 0$, left column) with the dynamics obtained with the band-restricted Hamiltonian (i.e., setting  $W=0$, right column) across different interaction regimes and initializations. Removing $W$ enforces strictly local, light-cone-like information spread in all cases shown. This shows that the W is the sole term responsible for nonlocal propagation.

\newpage 
\subsection{Role of Initial State ($z$ vs $x$ Basis)}
\label{zBasis}

\begin{figure}[htbp!]
    \centering
    \includegraphics[width=\linewidth]{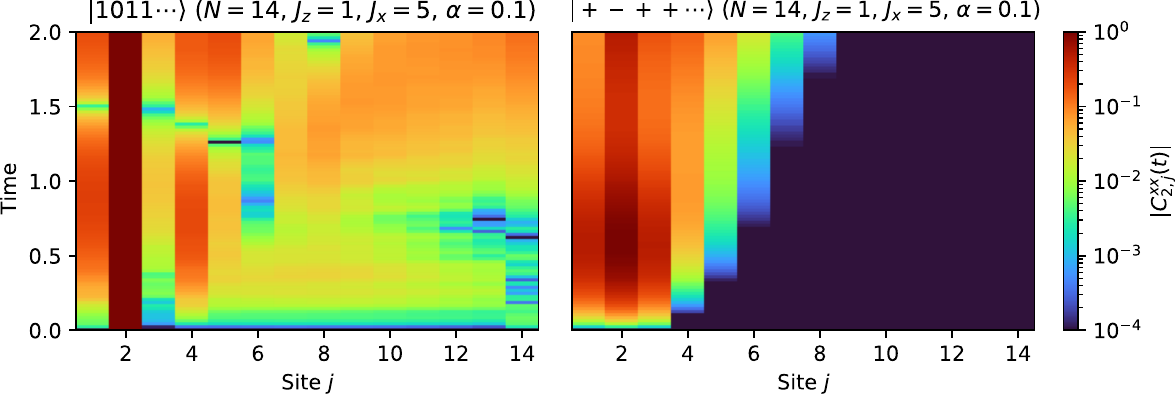}
    \caption{Basis dependence of information spread from a localized excitation. Time evolution of the connected correlator $|C_{2j}^{xx}(t)|$, Eq.~(\ref{eqsi:cor}) for two initial states containing a single excitation localized at site $i=2$. \textbf{Left:} Initial state prepared in the $z$ basis, for which the effective $W_{i,j}$ term couples all pairs of sites, leading to an immediate breakdown of locality and the emergence of nonlocal correlations. \textbf{Right:} Initial state prepared in the $x$ basis, where $W_{i,j}$ acts as a two-body interaction dependent on excitations in the $x$ basis, resulting in strictly local propagation throughout the bulk. Parameters are $N=14$, $J_z=1$, $J_x=5$, and $\alpha=0.1$. }
    \label{zinit}
\end{figure}

We show that the emergence of local or nonlocal information spread depends not only on the Hamiltonian, but also on the structure of the initial state.
In Figure~(\ref{zinit}) we consider the dynamics of the connected correlator for two different initial states, each containing a single localized excitation at site \(i=2\), but defined in different bases.

In the left panel, the initial state corresponds to a single spin flip relative to the fully polarized background, i.e., a state of the form \(|\downarrow  \uparrow_2 \downarrow \cdots \downarrow\rangle_z\)
where $z$ is the direction characterizing the nearest neighbor coupling.
In this basis, the effective term \(W_{i,j}\) acts as a long-range coupling between all site pairs, independent of the local excitation configuration. As a result, the dynamics immediately generate correlations across the entire chain, leading to a rapid breakdown of locality, and can be thought of as generating multiple nonlocal light cones emerging from all the sites.

In contrast, in the \(x\)-basis case (right panel), the initial state is  a single excitation relative to an \(x\)-polarized background, i.e., \(|\leftarrow \rightarrow_2\leftarrow  \cdots \leftarrow\rangle_x\)
where $x$ is the direction of the long-range Hamiltonian.
Here, the operator \(W_{i,j}\) takes the form of a density--density interaction in the \(x\) basis,  and therefore only contributes when multiple excitations are present. Since the initial state contains only a single excitation and $J_x$ is large, the \(W\) term makes a very small contribution, so that information spread is mainly local.

\subsection{Tunable Nonlocality}
\label{Tunable nonlocality}
We demonstrate that information propagation in our model requires the local $J_z$ hopping term, and that nonlocal light cones arise specifically from the interplay between this local coupling and the long-range interaction $W$. Furthermore, we show that this dependence allows for precise spatial control over signal propagation.

\begin{figure}[h!]
    \centering
    \includegraphics[width=\linewidth]{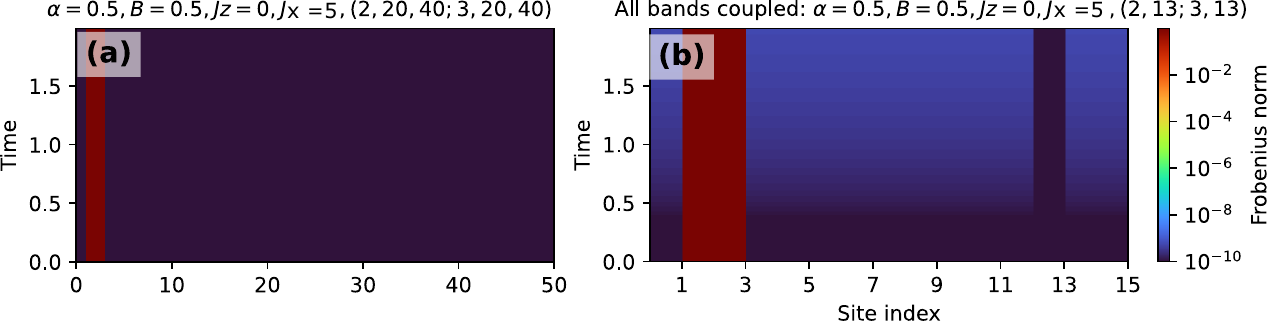}
    \caption{Density plots of the Frobenius norm showing light-cone dynamics under the band-3–projected Hamiltonian ${\hat H}_{\mathrm{eff}}^{b=3}$ for the initial states $(2,20,40;\,3,20,40)$ [Eq.~\eqref{pairnotation} in the main text] with $N=50$, $J_{\mathrm{x}}=0.5$, and $\alpha=0.5$.
    \textbf{(a)} $B$–$J_x$ model with an additional longitudinal field term $B \sum_i \sigma_i^z$ with $B=0.5$ and $J_z=0$.
    \textbf{(b)} Full Hamiltonian (without band projection) with N=15, initial states $(2,13;3,13)$ but with $J_z=0$ everywhere and a longitudinal $B$ field turned on, as in panel (a). These panels together confirm that the $W$ term alone cannot generate signal propagation; information spread  fundamentally requires the local coupling provided by $J_z$.
    }
    \label{BJlongPlots}
\end{figure}

The need for $J_z$ is first established by comparing projected and full dynamics. In Fig.~\ref{BJlongPlots}(a), we evolve the system under the band-projected Hamiltonian $\hat{H}_{\mathrm{eff}}^{b=3}$ with $J_z$ set to zero, but in the presence of a non-zero uniform magnetic field $B$ in the $z$-direction. Despite the presence of   $B$ and the long-range $W$ term, no light-cone structure appears. This is corroborated by the full Hamiltonian dynamics shown in Fig.~\ref{BJlongPlots}(b); even without band projection, the absence of $J_z$ results in a suppression of information spread for large $J_x$. These two panels together confirm that the $W$ term alone cannot generate signal propagation; information spread   fundamentally requires the local coupling provided by $J_z$.

\begin{figure}[h!]
    \centering
    \includegraphics[width=\linewidth]{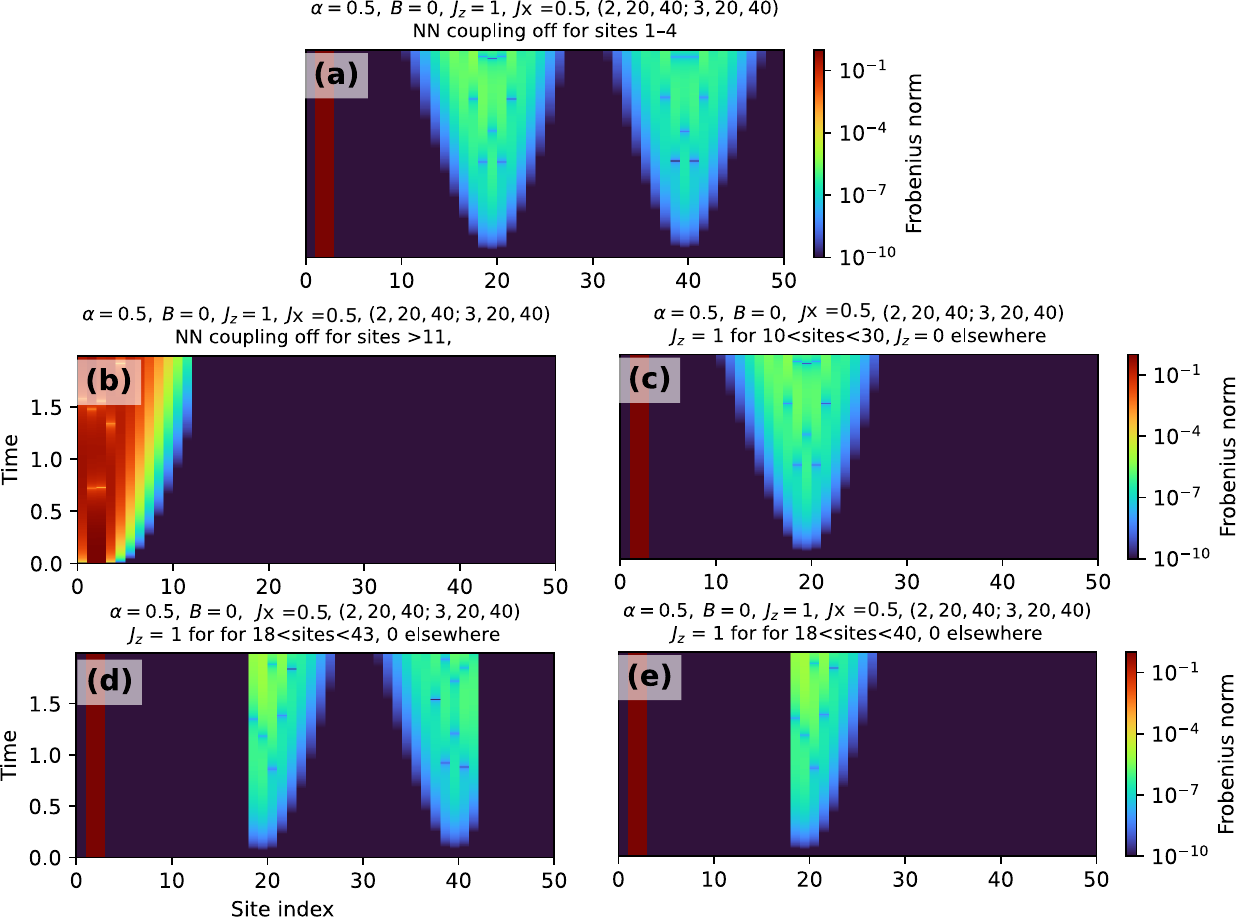}
    \caption{
    Density plots of the Frobenius norm showing light-cone dynamics under the band-3–projected Hamiltonian ${\hat H}_{\mathrm{eff}}^{b=3}$ for the initial states $(2,20,40;\,3,20,40)$ [Eq.~\eqref{pairnotation}  in the main text] with $N=50$, $J_{\mathrm{long}}=2$, and $\alpha=0.5$. Unlike in Fig. \ref{BJlongPlots}, here we set $B=0$
    \textbf{(a–e)} Hamiltonian with $J_z$ selectively disabled:
    \textbf{(a)} $J_z=0$ on sites 1–4 and $J_z=1$ elsewhere;
    \textbf{(b)} $J_z=0$ for sites $>11$ and $J_z=1$ elsewhere;
    \textbf{(c)} $J_z=1$ only in the region of the second light cone;
    \textbf{(d)} $J_z=1$ on sites 19–42;
    \textbf{(e)} same as (d) but excluding site 40.
    }
    \label{BJlongPlots2}
\end{figure}

The sensitivity of the nonlocal signal to the local hopping profile provides a mechanism for engineering information propagation. In Fig.~\ref{BJlongPlots2}(a–e), we selectively restore $J_z$ in specific spatial regions: Panels (a) and (b) show that disabling $J_z$ at the boundaries (sites 1–4 or sites $>11$) clips the light cone, confining the signal to regions where the local hopping is active. In panel (c), we restrict $J_z=1$ exclusively to the region surrounding the second excitation, which results in the emergence of only the corresponding nonlocal light cone, while the first remains suppressed. Panels (d) and (e) demonstrate fine-grained control; by activating $J_z$ only on sites 19–42 (panel d) and then specifically excluding a single site (site 40 in panel e), we show that the nonlocal signal can be masked at will. These results establish that nonlocal light cones originate from the synergy between the long-range $W$ interaction and $J_z$ hopping. By shaping the local profile of $J_z$, the presence, speed, and location of the nonlocal signal can be tuned. Consequently, local modulation of $J_z$ provides a practical and experimentally accessible mechanism to engineer and control nonlocal information transport in long-range systems.

\subsection{2D Extension}
\label{2DIsing}
\begin{figure}[htbp!]
    \centering
    \includegraphics[width=\linewidth]{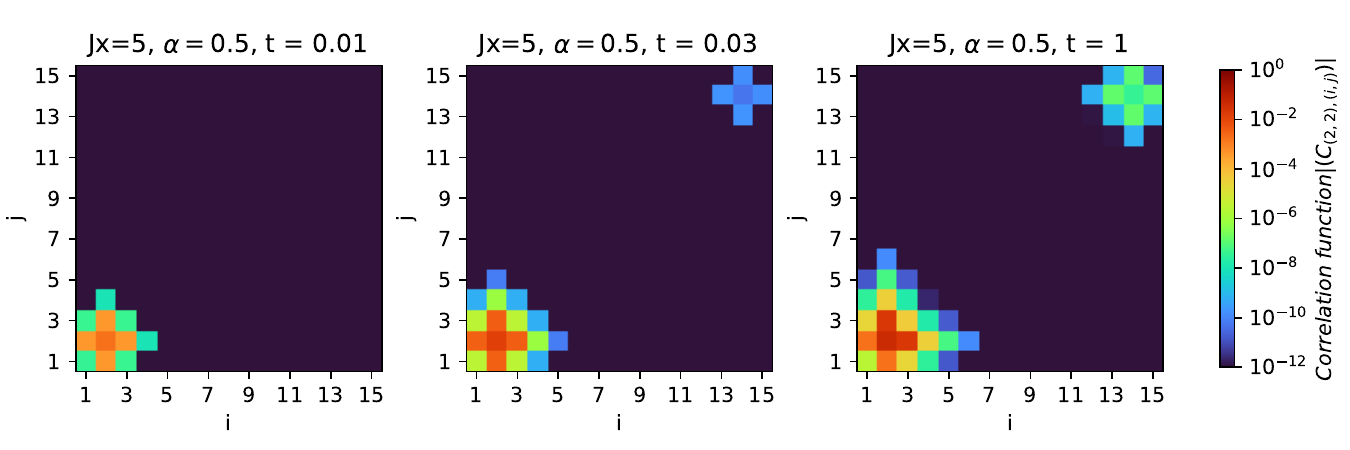}
    \caption{Light cones on a 2D lattice. 
    The connected two-point correlator, $|\langle \sigma_{2,2}^x(t)\sigma_{i,j}^x(t)\rangle
    - \langle \sigma_{2,2}^x(t)\rangle \langle \sigma_{i,j}^x(t)\rangle|$, is computed for the projected Hamiltonian $\hat H_{\mathrm{eff}}^{b=2}$ at fixed time slices, starting with 
     a band-2 initial state with excitations at site (2,2) and (14,14). The panels show nonlocal correlations developing between the excitations, demonstrating that correlations can be programmed using a 2D lattice Hamiltonian.}
    \label{2dIsingN10Jx5}
\end{figure}

We extend the model to a two-dimensional square lattice with the Hamiltonian
\begin{equation}
\hat H =
J_z \sum_{\langle \mathbf{r},\mathbf{r}' \rangle}
\sigma_{\mathbf{r}}^z \sigma_{\mathbf{r}'}^z
+
J_x
\sum_{\mathbf{r}<\mathbf{r}'}
\frac{1}{|\mathbf{r}-\mathbf{r}'|^\alpha}
\,\sigma_{\mathbf{r}}^x \sigma_{\mathbf{r}'}^x ,
\end{equation}
where \(\mathbf{r}=(i,j)\). In the $J_z$ term, $\langle \mathbf{r},\mathbf{r}' \rangle$ denotes nearest-neighbor coupling, i.e., $\mathbf{r}=(i,j)$ is coupled to  nearest neighbors \(\mathbf{r}'=(i\pm1,j),\,(i,j\pm1)\). Projecting onto a fixed band $b$ yields \(\hat H_{\mathrm{eff}}^{b}\).

Figure~\ref{2dIsingN10Jx5} shows dynamics under \(\hat H_{\mathrm{eff}}^{b=2}\) for two excitations at \((2,2)\) and \((14,14)\), using the connected correlator
\[
\left|
\langle \sigma_{2,2}^x(t)\sigma_{i,j}^x(t)\rangle
-
\langle \sigma_{2,2}^x(t)\rangle
\langle \sigma_{i,j}^x(t)\rangle
\right|.
\]
Correlations develop nonlocally between distant regions as time increases from left to right panels, demonstrating programmable nonlocal light cones also in 2D systems.

\section{Derivation of the Intraband Signal}
\label{intraband scaling}

\subsection{Model and Hamiltonians}

In the regime where interband transitions (changing the total boson number $b$) are strongly suppressed, we neglect the pair creation and annihilation terms in Eq.~\eqref{effham0} in the main text. For example, if we restrict the dynamics strictly to the $b=2$ intraband block, the resulting intraband Hamiltonian is:
\begin{align}
\label{intrabandEffHAM}
\hat{H}_{\mathrm{intraband}}^{b=2} &= J_z \sum_i \left( \hat{a}_i^\dagger \hat{a}_{i-1} + \hat{a}_{i-1}^\dagger \hat{a}_i \right) \nonumber \\
&+ \sum_i V(i) \,\hat{n}_i + \sum_{i<j} \frac{4 J_x}{|i-j|^\alpha} \,\hat{n}_i \hat{n}_j \,.
\end{align}

When the system is placed on a ring with periodic boundary conditions (PBC), the lattice becomes translationally invariant. Thus, the effective local potential $V(i)$ becomes
\begin{equation}
V(i) = -2 J_x \sum_{j \ne i} \frac{1}{|i-j|^\alpha} \equiv V_{\mathrm{const}} \,,
\end{equation}
independent of the specific site $i$. Furthermore, since $\hat{H}_{\mathrm{intraband}}$ strictly conserves the total particle number $b = \sum_i \hat{n}_i$, the potential term simplifies to $V_{\mathrm{const}} \sum_i \hat{n}_i = b V_{\mathrm{const}}$. This merely introduces a global energy shift to the specific $b$-particle sector. It commutes with the rest of the Hamiltonian and does not contribute to the quantum dynamics within that band.

We are left in Eq.~\eqref{intrabandEffHAM} with a nearest-neighbor hopping term and a two-body potential.

\subsection{Frobenius Norm as the Boson Occupation Difference}
We measure the effect of starting in two different initial states, $\ket{\psi_1}$ and $\ket{\psi_2}$, using the normalized Frobenius norm of the difference $\Delta\rho_n=\rho^{(1)}_n-\rho^{(2)}_n$ between the corresponding local reduced density matrices $\rho^{(1)}_n$ and $\rho^{(2)}_n$, at site $n$:
\begin{equation}
\|\Delta \rho_n(t)\|_F = \sqrt{\text{Tr}(\Delta\rho_n \Delta\rho_n^\dagger)} \,.
\end{equation}
For hard-core bosons, the local density matrix is a $2 \times 2$ matrix.
Because the intraband Hamiltonian strictly conserves the particle number (a $U(1)$ symmetry), local off-diagonal coherences such as $\langle \hat{a}_n \rangle$ and $\langle \hat{a}_n^\dagger \rangle$ are identically zero for initial conditions with a fixed number of bosons.

Thus, the local density matrices are entirely diagonal: $\rho_n = \text{diag}(1-\langle \hat{n}_n \rangle, \langle \hat{n}_n \rangle)$. The difference is $\Delta \rho_n = \text{diag}(-\Delta n_n, \Delta n_n)$, where $\Delta n_n = \langle \hat{n}_n^{(1)} \rangle - \langle \hat{n}_n^{(2)} \rangle$. Plugging this into the definition:
\begin{equation}
\|\Delta \rho_n(t)\|_F = \sqrt{(-\Delta n_n)^2 + (\Delta n_n)^2} = \sqrt{2} |\Delta n_n(t)| \,.
\end{equation}
After rescaling by the maximum possible norm ($\sqrt{2}$), the normalized Frobenius norm is exactly equal to the absolute difference in local boson densities: $|\langle \hat{n}_n^{(1)}(t) \rangle - \langle \hat{n}_n^{(2)}(t) \rangle|$.

\subsection{Time Evolution and Short-Time Expansion}

In the following, we present the calculation for the $b=2$ case to simplify the notation, but the argument applies generally to bands $b \ge 2$. We consider two hard-core bosons initialized on a one-dimensional lattice at two distant sites $m$ and $q$:
\begin{equation}
    |\Psi_m\rangle = |m,q\rangle\,,
\end{equation}
where the particle at site $m$ acts as the ``message'' particle and the particle at site $q$ acts as the ``target.'' The intraband Hamiltonian takes the form of a sum of a diagonal term and a hopping term,
\begin{equation}
    H_{\mathrm{intraband}} \equiv H = D + T\,,
\end{equation}
with the $D$ term containing the diagonal interaction energies $W_{i,j} = \frac{4J_x}{|i-j|^\alpha}$ and the $T$ term containing nearest-neighbor hopping with amplitude $J_z$.

We probe information transfer by comparing the occupation dynamics at site $q$ for two different initial message locations, $m=2$ and $m=3$. The intraband signal is therefore defined as
\begin{equation}
    \mathcal{F}_{\mathrm{intraband}}(t)=|\langle \hat n_{q}(t)\rangle_{m=3}-\langle \hat n_{q}(t)\rangle_{m=2}| \,.
\end{equation}

The time-evolved density operator is
\begin{equation}
    \hat n_{q}(t) = U^\dagger(t)\hat n_{q}U(t), \qquad U(t)=e^{-iHt}.
\end{equation}

Expanding the evolution operators in powers of time gives
\begin{align}
U(t)
&=
\mathbb{I}
-iHt
-\frac{t^2}{2}H^2
+\frac{it^3}{6}H^3
+\frac{t^4}{24}H^4
+\mathcal{O}(t^5),
\\
U^\dagger(t)
&=
\mathbb{I}
+iHt
-\frac{t^2}{2}H^2
-\frac{it^3}{6}H^3
+\frac{t^4}{24}H^4
+\mathcal{O}(t^5).
\end{align}

The occupation expectation value
\begin{equation}
c_m(t)
=
\langle \Psi_m|
U^\dagger(t)\hat n_{q}U(t)
|\Psi_m\rangle
\end{equation}
can then be organized as a short-time series:
\begin{equation}
c_m(t)
=
\sum_{k=0}^{\infty}
t^k c_m^{(k)}
=
c_m^{(0)}
+
t\,c_m^{(1)}
+
t^2 c_m^{(2)}
+
t^3 c_m^{(3)}
+
t^4 c_m^{(4)}
+
\mathcal{O}(t^5).
\end{equation}

Since the normalized Frobenius norm reduces exactly to the absolute difference in local densities, the intraband signal is determined by the differences between the short-time expansion coefficients,
\begin{equation}
\Delta c^{(k)}
=
c_3^{(k)}
-
c_2^{(k)}.
\end{equation}
The leading nonvanishing coefficient therefore determines the earliest-order contribution to the signal, or more explicitly,
\begin{equation}
    \mathcal{F}_{\text{intraband}}(t) =  t^k \Delta c^{(k)} + \mathcal{O}(t^{k+1}) \,,
\end{equation}
where $k$ is the first nonvanishing order.

\subsection{Operator Path Strings and Partitioning}

Instead of evaluating each perturbative coefficient $c_m^{(k)}$ separately, it is useful to organize the entire short-time expansion in terms of operator strings. This provides a transparent way to classify which virtual processes contribute to the signal and which are forbidden by constraints.

Expanding the evolution operators in powers of time generates products of $H$ acting on both sides of the number operator $\hat n_{q}$. At perturbative order $k$, the generic contribution takes the form
\begin{equation}
\langle \Psi_m|
H^{k-i}\,\hat n_{q}\,H^i
|\Psi_m\rangle,
\qquad
i=0,\dots,k.
\end{equation}

Since each factor of $H$ can be either $D$ or $T$, every term can be represented as a binary operator string composed of diagonal operators and hopping operators. For example, $DTTT$ represents the ordered sequence $T \rightarrow T \rightarrow T \rightarrow D.$

The insertion of $\hat n_{q}$ naturally partitions each string into two segments,
\begin{equation}
\underbrace{
\langle \Psi_m|
\hat O_k \cdots \hat O_{i+1}
}_{\text{left segment}}
\,
\hat n_{q}
\,
\underbrace{
\hat O_i \cdots \hat O_1
|\Psi_m\rangle
}_{\text{right segment}},
\end{equation}
where each operator $\hat O_j \in \{D,T\}$.\\

The right segment originates from the forward-time evolution operator $U(t)$ and evolves the initial state $|\Psi_m\rangle$ into a superposition of intermediate configurations. The number operator $\hat n_{q}$ then acts as a local measurement cut through the operator string, selecting only those intermediate states in which site $q$ is occupied. Finally, the left segment, originating from $U^\dagger(t)$, evolves the intermediate configuration back toward the final state $\langle\Psi_m|$.

For example, a term from the fourth-order contribution
\begin{equation}
\langle H^2 \hat n_{q} H^2\rangle
\end{equation}
contains all $2^4 = 16$ possible four-operator strings with the cut fixed in the middle of the string:
\begin{align}
&DD|DD,\quad DD|DT,\quad DD|TD,\quad DD|TT, \nonumber\\
&DT|DD,\quad DT|DT,\quad DT|TD,\quad DT|TT, \nonumber\\
&TD|DD,\quad TD|DT,\quad TD|TD,\quad TD|TT, \nonumber\\
&TT|DD,\quad TT|DT,\quad TT|TD,\quad TT|TT.
\end{align}

Each operator string corresponds to a virtual quantum trajectory through Hilbert space, while the position of the cut (vertical bar) specifies the point at which the occupation at site $q$ is probed by $\hat n_{q}$.

\subsection{Path Constraints and Reduction of Operator Strings (Selection Rules)}

Although the perturbation series formally contains exponentially many operator strings, most virtual trajectories are eliminated by some constraints. Introducing these constraints progressively allows the structure of the short-time expansion to emerge naturally.

\vspace{0.5em}

We recall that the initial two-particle configuration is denoted as
\begin{equation}
|m,q\rangle \,,
\end{equation}
where $m$ labels the message particle and $q$ labels the target particle. The hopping operator $T$ moves exactly one particle by one lattice site:
\begin{equation}
|m,q\rangle
\;\xrightarrow{\,T\,}\;
|m\pm1,q\rangle
\quad \text{or} \quad
|m,q\pm1\rangle \,.
\end{equation}

The diagonal operator $D$ acts locally by multiplying a configuration by its interaction energy, without changing the occupation pattern:
\begin{equation}
D|m,q\rangle = W_{m,q}\,|m,q\rangle.
\end{equation}

\subsubsection*{Constraint 1: Closed-Loop Constraint}

Every perturbative coefficient appears inside an expectation value of the form
\begin{equation}
\langle m,q| \cdots |m,q\rangle.
\end{equation}

Therefore, any nonzero virtual trajectory must eventually return to the original configuration $\langle m,q|$. Since each application of $T$ changes the message or the particle configuration, a necessary condition for a nonvanishing contribution is that the total number of hopping operators be even. All strings containing an odd number of $T$ operators vanish identically, since they cannot return the system to its initial configuration.

\subsubsection*{Constraint 2: Measurement-Cut Hole Constraint}

The key simplification comes from rewriting the measurement operator as
\begin{equation}
\hat{n}_{q} = \mathbb{I} - \bar{n}_{q},
\end{equation}
where $\bar{n}_{q}$ projects onto configurations in which site $q$ is empty. Substituting this decomposition into the general $k$-th order coefficient,
\begin{equation}
c_m^{(k)}
=
\sum_{j=0}^{k}
\lambda_j^{(k)}
\left\langle
H^{k-j}\,
\hat n_{q}\,
H^j
\right\rangle,
\end{equation}
gives us
\begin{equation}
c_m^{(k)}
=\sum_{j=0}^{k}
\lambda_j^{(k)}
\langle H^k\rangle-
\sum_{j=0}^{k}
\lambda_j^{(k)}
\left\langle
H^{k-j}\,
\bar n_{q}\,
H^j
\right\rangle \,.
\end{equation}

The coefficients $\lambda_j^{(k)}$ arise from the forward-backward Taylor expansion of $e^{iHt}\hat n_{q}e^{-iHt}.$ At order $k$, they take the general binomial form
\begin{equation}
\lambda_j^{(k)}
=
\frac{1}{k!}
(-1)^j
\binom{k}{j} \,.
\end{equation}
Using the binomial theorem to expand $(x + y)^k$,
$$(x + y)^k = \sum_{j=0}^{k} \binom{k}{j} x^{k-j} y^j \,,$$
and setting $x = 1$ and $y = -1$, the expansion becomes:
$$(1 - 1)^k = \sum_{j=0}^{k} (-1)^j \binom{k}{j} = 0\,.$$
For any perturbative order $k \geq 1$,
\begin{equation}
    \sum_{j=0}^{k}
\lambda_j^{(k)}
\langle H^k\rangle = \langle H^k\rangle\sum_{j=0}^{k}
\lambda_j^{(k)}= 0 \,.
\end{equation}
Therefore,
\begin{equation}
c_m^{(k)}
=
-
\sum_{j=0}^{k}
\lambda_j^{(k)}
\left\langle
H^{k-j}\,
\bar n_{q}\,
H^j
\right\rangle.
\end{equation}

The signal is therefore entirely determined by the hole operator $\bar n_{q}$. This operator imposes the constraint that site $q$ must be empty at the measurement cut, which places three conditions on any contributing trajectory:
\begin{itemize}
    \item[] (1) the target particle must hop away from site $q$ before the cut;
    \item[] (2) the target particle must remain away from site $q$ when the cut is applied;
    \item[] (3) the target particle must return afterward in order to satisfy the closed-loop condition of Constraint 1.
\end{itemize}

These requirements immediately eliminate large classes of operator strings.

\textbf{Constraint 2a:} Purely D strings containing only $D$ operators can never satisfy condition (1), since $D$ does not move the target particle. Consequently, all such strings are annihilated by $\bar n_{q}$ no matter where the cut is placed.

\textbf{Constraint 2b:} Conditions (1) and (3) dictate that any surviving trajectory must contain at least two hopping operators $T$: one to move the target particle away from site $q$ before the measurement cut, and another to return it afterward. Generally, each $T$ operator acts on either the message ($m$) or target ($q$) particle degree of freedom:
\begin{align}
|m,q\rangle &\to |m\pm1,q\rangle \to |m,q\rangle, \nonumber \\|m,q\rangle &\to |m,q\pm1\rangle \to |m,q\rangle.\label{Thopping}
\end{align}

While evaluating the signal naively requires summing over all such configurations, Constraint 2b simplifies the summation. If a trajectory utilizes its two available $T$ operators exclusively for message-particle hops, the target particle remains stationary at site $q$, violating condition (1). Consequently, for all operator strings containing exactly two $T$ operators, both hops are strictly constrained to the target particle. The only non-vanishing processes are:
\begin{equation}
|m,q\rangle \to |m,q\pm1\rangle \to |m,q\rangle.
\end{equation}

\textbf{Constraint 2c:} The hole projector requires site $q$ to be empty at the partition, so the target particle must hop away from site $q$ before the cut and return afterward. This immediately eliminates any partition in which both hopping operators occur entirely on the same side of the cut. If an even number of $T$ operators appear before the cut, the target necessarily performs an immediate round trip and reoccupies site $q$ before the projector acts. Likewise, if an even number of hopping operators occur after the cut, the target has not yet left site $q$ when the projector is evaluated. Both cases violate the vacancy condition imposed by $\bar n_{q}$. Consequently, the only surviving partitions are those containing an odd number of $T$ operators on each side of the cut, while the total number of $T$ operators in the string is even, consistent with Constraint 1.

\subsubsection*{Constraint 3: Purely $T$ Strings}

Operator strings consisting only of hopping operators $T$ do not contribute to the intraband signal. These trajectories generate purely kinetic motion with no insertion of diagonal interaction energies $D$, and therefore evolve identically for both initial configurations $|3,q\rangle$ and $|2,q\rangle$. Since all message dependence enters exclusively through the interaction energies $W_{m,q}$ encoded in $D$, purely-$T$ strings are independent of $m$.

Formally, all strings of the form $T^k$ contribute equally to $c_3^{(k)}$ and $c_2^{(k)}$, and hence cancel exactly in the difference
\begin{equation}
\Delta c^{(k)} = c_3^{(k)} - c_2^{(k)}.
\end{equation}

This removes all purely kinetic trajectories, leaving only mixed $D$--$T$ processes in which the target particle acquires interaction-dependent phases during its virtual excursions.

\subsection{Vanishing of Lower Orders and Emergence of the Fourth-Order Signal}
\textbf{Zeroth order ($t^0$):}
At zeroth order, no hopping processes are available. Since Constraint 2 requires the target particle to leave site $q$ before the measurement cut, the hole condition can never be satisfied. Therefore, all zeroth-order contributions vanish in the signal. Consequently,
\begin{equation}
c_m^{(0)} = 0.
\end{equation}

\textbf{First order ($t^1$):}
At first order, the only possible strings are $D$ and $T$. Purely ($D$) paths are eliminated by Constraint 2a, while a single hopping process ($T$) cannot satisfy Constraint 1 and also Constraint 3. Consequently,
\begin{equation}
c_m^{(1)} = 0.
\end{equation}

\textbf{Second order ($t^2$):}
At second order, Constraint 1 permits only $TT$ or $DD$ strings. The purely diagonal string $DD$ is eliminated by Constraint 2a, while two-hop $TT$ strings are eliminated by Constraint 3, consequently,
\begin{equation}
c_m^{(2)} = 0.
\end{equation}

\textbf{Third order ($t^3$):}
Applying Constraint 2, the expansion produces: \begin{align}
\label{c3hole}
c_m^{(3)}
&=
-\frac{1}{6}\langle H^3 \bar n_q \rangle
+
\frac{1}{2}\langle H^2 \bar n_q H \rangle
-
\frac{1}{2}\langle H \bar n_q H^2 \rangle
+
\frac{1}{6}\langle \bar n_q H^3 \rangle \,.
\end{align}

Constraint 1 requires an even number of hopping operators $T$. At third order, the only allowed operator strings are the purely diagonal string $DDD$ and permutations of $TTD$:
\begin{align}
TTD,\qquad TDT,\qquad DTT.
\end{align}

The $DDD$ contribution is eliminated by Constraint 2a. Applying Constraint 2c to the remaining strings leaves the partitions
\begin{align}
T|TD,\qquad T|DT,\qquad TD|T,\qquad DT|T .
\end{align}

The surviving partitions inherit their weights directly from the Taylor coefficients multiplying each partition structure in Eq.~\eqref{c3hole}. The partitions with the $D$ operator to the left of the cut have coefficient $\frac{1}{2}$, while the ones where the $D$ operator is on the right have coefficient $-\frac{1}{2}$.

When the terms are added up with these Taylor coefficient weights, the sum vanishes. Consequently,
\begin{equation}
c_m^{(3)} = 0.
\end{equation}

\textbf{Fourth-Order ($t^4$):}

At fourth order, applying Constraint 2, the expansion produces:
\begin{align}
\label{c4hole}
c_m^{(4)}
&=
-\frac{1}{24}\langle H^4 \bar n_{q} \rangle
+
\frac{1}{6}\langle H^3 \bar n_{q} H \rangle
\nonumber\\
&\quad
-
\frac{1}{4}\langle H^2 \bar n_{q} H^2 \rangle
+
\frac{1}{6}\langle H \bar n_{q} H^3 \rangle
-
\frac{1}{24}\langle \bar n_{q} H^4 \rangle .
\end{align}

Expanding $H = D + T$ generates $2^4 = 16$ operator strings, with 5 possible placements of the cut, giving 80 formal paths. Each of these further branches into multiple microscopic trajectories, since every hopping operator $T$ may act on either the message ($m$) or target ($q$) particle and in either direction, leading to up to $4^4 = 256$ microscopic trajectories per operator string. This results in an unwieldy number of contributions, which is precisely why the constraints defined earlier are useful. The 16 unpartitioned strings are
\begin{align}
&DDDD,\quad DDDT,\quad DDTD,\quad DDTT,\nonumber\\
&DTDD,\quad DTDT,\quad DTTD,\quad DTTT,\nonumber\\
&TDDD,\quad TDDT,\quad TDTD,\quad TDTT,\nonumber\\
&TTDD,\quad TTDT,\quad TTTD,\quad TTTT.
\end{align}

Applying Constraints 1, 2, 2a, and 3 reduces this set to the 6 strings
\begin{align}
&TTDD,\quad TDTD,\quad TDDT,\nonumber\\
&DTTD,\quad DTDT,\quad DDTT, 
\end{align}

Constraint 2b now fixes the message particle, since we only have two $T$ operators available and both go into moving the target particle. At this stage, each of the 6 strings admits 5 possible placements of the measurement cut, giving 30 partitioned contributions in total. Each partition still contains $2^2 = 4$ microscopic hopping histories, corresponding to the choice of whether each $T$ acts on the left or right target hop direction. Therefore, the full fourth-order sector contains up to $30 \times 4 = 120$ microscopic trajectories.

At this stage, Constraint 2c further reduces the cuts to:
\begin{align}
&T|TDD,\quad
T|DTD,\quad
TD|TD,\quad
T|DDT,\quad
TD|DT,\nonumber\\
&TDD|T,\quad
DT|DT,\quad
DTD|T,\quad 
DT|TD,\quad
DDT|T \,.
\end{align}

The surviving partitions inherit their weights directly from the Taylor coefficients multiplying each partition structure in Eq.~\eqref{c4hole} and can be classified as symmetric and asymmetric based on the placement of the two $D$ operators relative to the cut. The symmetric partitions, with a $D$ operator on either side of the cut, have coefficients $-\frac{1}{4}$, while the asymmetric ones have coefficient $+\frac{1}{6}$:

\begin{align}
&+\frac{1}{6}\,T|TDD,\quad
+\frac{1}{6}\,T|DTD,\nonumber\\
&-\frac{1}{4}\,TD|TD,\quad
+\frac{1}{6}\,T|DDT,\nonumber\\
&-\frac{1}{4}\,TD|DT,\quad
+\frac{1}{6}\,TDD|T,\nonumber\\
&-\frac{1}{4}\,DT|DT,\quad
+\frac{1}{6}\,DTD|T,\nonumber\\
&-\frac{1}{4}\,DT|TD,\quad
+\frac{1}{6}\,DDT|T.
\end{align}

At this point, the partitions have already served their purpose of assigning the correct Taylor weights. Since all surviving terms correspond to the same underlying target-particle excursions, the cut placement no longer produces distinct contributions. We can therefore discard the partition labels and regroup the terms by identical operator content, keeping only the effective surviving structure.

The initial state is $|m,q\rangle$ with energy $W_q \equiv W_{m,q}$, and the intermediate energy is $W_s \equiv W_{m,s}$ for $s=q \pm 1$. For each of the 6 strings, we get:

$$TTDD \to +\frac{1}{6} J_z^2 \sum_s W_q^2, \quad TDTD \to -\frac{1}{12} J_z^2 \sum_s W_qW_s,$$
$$TDDT \to +\frac{1}{12} J_z^2 \sum_s W_s^2, \quad DTDT \to -\frac{1}{12} J_z^2 \sum_s W_qW_s,$$
$$DTTD \to -\frac{1}{4} J_z^2 \sum_s W_q^2, \quad DDTT \to +\frac{1}{6} J_z^2 \sum_s W_q^2.$$

Combining these, we obtain the total coefficient $c_m^{(4)}$,
$$c_m^{(4)} = J_z^2 \sum_s \left( \frac{1}{12}W_q^2 - \frac{1}{6}W_qW_s + \frac{1}{12}W_s^2 \right) \,,$$
or
$$c_m^{(4)} = \frac{J_z^2}{12} \sum_{s\in\{q-1,q+1\}} (W_q - W_s)^2 \,.$$

The leading-order intraband signal is then
\begin{align}
\Delta c_m^{(4)}
=
c_3^{(4)}-c_2^{(4)} \,.
\end{align}

To determine the asymptotic scaling at large separation, we define
\begin{equation}
r_0 = q-m\,,
\qquad
x = \frac{1}{r_0} \ll 1.
\end{equation}

The interaction energies at neighboring sites are expanded using
\begin{equation}
(1\pm x)^{-\alpha}
\approx
1
\mp
\alpha x
+
\frac{\alpha(\alpha+1)}{2}x^2.
\end{equation}

Since
\begin{equation}
W_{m,s}
=
\frac{4J_x}{|m-s|^\alpha},
\end{equation}
the neighboring interaction differences become
\begin{equation}
W_q-W_{q \pm 1}
\approx
4J_x\,R^{-\alpha}
\left[
\mp \alpha x
-
\frac{\alpha(\alpha+1)}{2}x^2
\right].
\end{equation}
Squaring and summing the contributions from sites $s=q \pm 1$ yields
\begin{align}
S(r_0)
&\equiv
\sum_{s=q \pm 1}
(W_q-W_s)^2
\nonumber\\
&\approx
2\alpha^2(4J_x)^2r_0^{-(2\alpha+2)}.
\end{align}

The discrete signal difference between neighboring message-particle positions is well approximated by a spatial derivative:
\begin{equation}
\Delta S
=
S(r_0-1)-S(r_0)
\approx
-
\frac{\partial S}{\partial r_0}.
\end{equation}
Differentiating gives
\begin{equation}
-
\frac{\partial S}{\partial r_0}
=
2\alpha^2(2\alpha+2)(4J_x)^2
r_0^{-(2\alpha+3)}\,,
\end{equation}
and substituting this back into the fourth-order expression gives the asymptotic intraband signal
\begin{align}
\label{Intraband_scaling_SI}
\mathcal{F}_{\mathrm{intraband}}^{(4)}
=
\frac{t^4J_z^2(4J_x)^2}{3}
\alpha^2(\alpha+1)
\frac{1}{r_0^{2\alpha+3}} + \mathcal{O}(t^5) \,.
\end{align}

\subsection{Derivation of the Optimal Interaction Exponent $\alpha_{\mathrm{max}}$}
To find the interaction exponent $\alpha$ that maximizes the long-range signal at a fixed macroscopic distance $r_{0}$, we maximize the $\alpha$-dependent portion of the asymptotic formula:
\begin{equation}
g(\alpha)=\frac{\alpha^{3}+\alpha^{2}}{r_{0}^{2\alpha}} \,.
\end{equation}
Setting the derivative $g^{\prime}(\alpha)=0$ yields
\begin{equation}
\frac{3\alpha^{2}+2\alpha}{r_{0}^{2\alpha}}-\frac{2 \ln(r_{0})(\alpha^{3}+\alpha^{2})}{r_{0}^{2\alpha}}=0 \,,
\end{equation}
and factoring out $\alpha/{r_{0}}^{2\alpha}$ (for $\alpha>0$) leaves the quadratic equation
\begin{equation}
2 \ln(r_{0})\alpha^{2}+(2 \ln(r_{0})-3)\alpha-2=0 \,.
\end{equation}
Solving for the positive root gives the exact optimal $\alpha$:
\begin{equation}
\label{alphamaxexact}
\alpha_{\mathrm{max}}=\frac{3-2 \ln(r_{0})+\sqrt{4 \ln^{2}(r_{0})+4 \ln(r_{0})+9}}{4 \ln(r_{0})} \,.
\end{equation}
In the limit of very large lattices $(r_{0}\rightarrow\infty)$, we expand the square root as $\sqrt{4 \ln^{2}(r_{0})+\dots} \approx 2 \ln(r_{0})$. 
Substituting this into the numerator yields the asymptotic scaling:
\begin{equation}
\label{alphamax}
\alpha_{\mathrm{max}}\approx\frac{1}{\ln(r_{0})} \,.
\end{equation}
Thus, as distance approaches infinity, the optimal exponent approaches the all-to-all limit $(\alpha\rightarrow 0)$.

\subsection{Analysis of Arrival of Nonlocal Signals }

Here, we provide a quantitative analysis of the arrival times of the programmable nonlocal signal.

We define the arrival time $t_{\mathrm{arr}}$ as the time required for the nonlocal signal at a target site $r$ to reach a predefined detection threshold $\theta$. As discussed in Eq.~(\ref{Intraband_scaling_SI}), the leading-order short-time contribution to the intraband signal, measured via the Frobenius norm, is given by:
\begin{equation}
||\Delta\rho_n(t)||_F^{\text{intraband}} \approx \frac{16}{3} t^4 J_z^2 J_x^2 \alpha^2 (\alpha+1) \frac{1}{r^{2\alpha+3}}
\end{equation}
Setting this expression equal to the threshold $\theta$ and solving for $t_{\mathrm{arr}}$, we obtain the explicit arrival time for the nonlocal channel:
\begin{equation}
\label{tarr}
t_{\mathrm{arr}} \approx \left( \frac{3 \theta r^{2\alpha+3}}{16 J_z^2 J_x^2 \alpha^2 (\alpha+1)} \right)^{\frac{1}{4}}
\end{equation}
Notably, this time exhibits an algebraic scaling with distance, $t_{\mathrm{arr}} \propto r^{\frac{2\alpha+3}{4}}$. For interaction exponents $\alpha < 1/2$, this results in a faster-than-ballistic propagation.

\subsection{Numerical Study of the Intraband Signal} Here we test our analytical results for the intraband signal numerically, specifically checking the spatial dependence and the dependence on $J_x$, $N$, $\alpha$, and time.



\begin{figure}[htbp!]
    \centering
    \includegraphics[width=\linewidth]{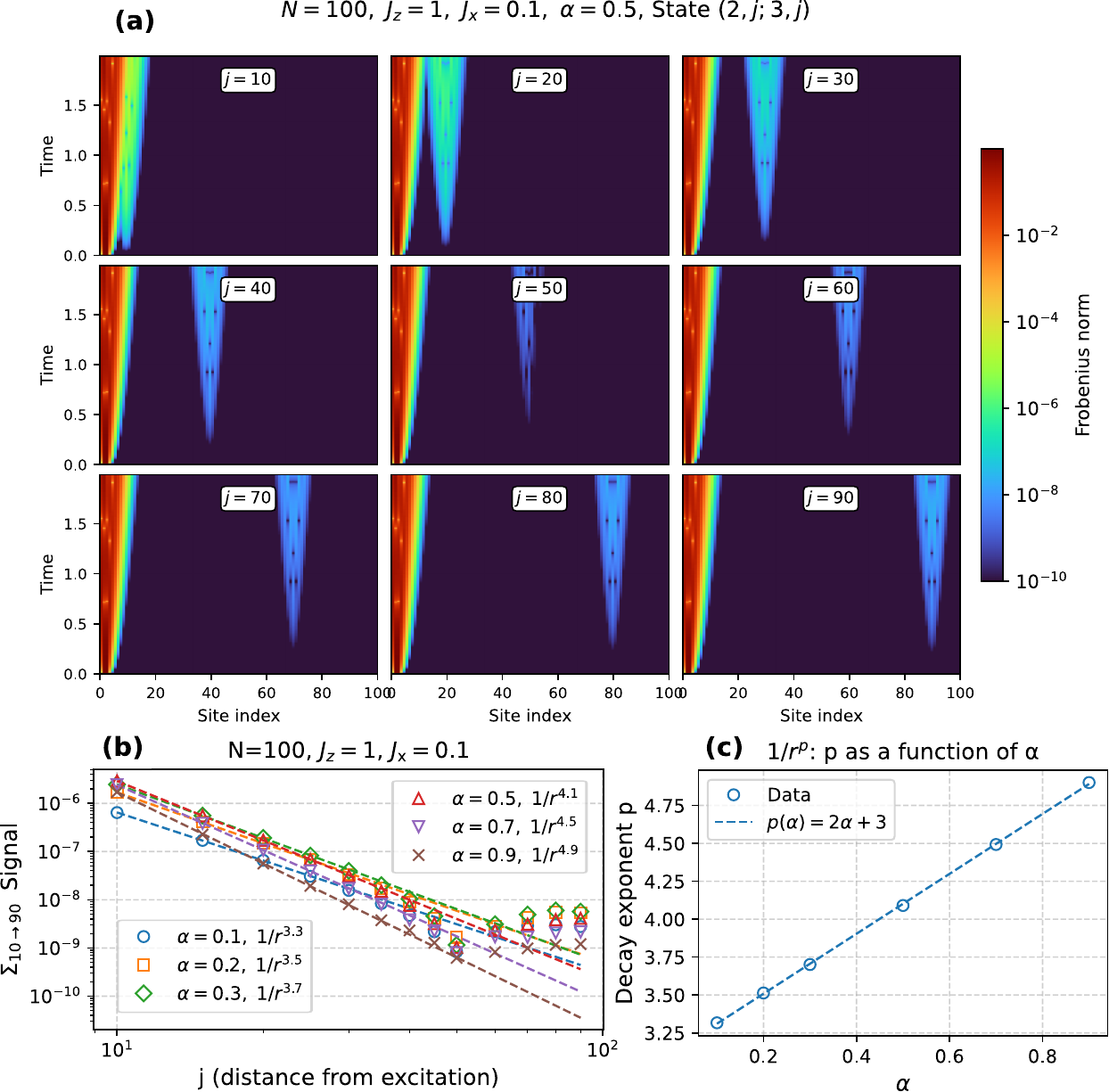}
    \caption{
   \textbf{(a)} The nine density plots show the Frobenius norm for initial states $(2,j;\,3,j)$, for nine different excitation positions $j$, evolved by the band-2 projected Hamiltonian ${\hat H}_{\mathrm{eff}}^{b=2}$. \textbf{(b)} shows the Frobenius norm signal integrated from site $10$ to site $90$ at $t=0.5$ versus initial excitation position $j$, for $N=100$, $J_x=0.1$, and $\alpha=0.5$. The dashed lines show $1/r^{p(\alpha)}$ scaling of the signal, with $r=j-3$, up to mid-chain, beyond which boundary effects appear for open boundary conditions.
    \textbf{(c)} shows the exponent $p(\alpha)$ for different $\alpha$ values. The exponent increases as $\approx 3+2 \alpha$, as predicted by Eq.~\eqref{Intraband_scaling_SI}.}
    \label{N100LFIMb2lcfara05VaryPosOBC}
\end{figure}

\textbf{Spatial Dependence:} We examine the spatial dependence by varying the initial excitation position $j$ (corresponding to the initial state pair $(2,j;3,j)$) under open boundary conditions. The top nine density plots in Fig. \ref{N100LFIMb2lcfara05VaryPosOBC} demonstrate that while the nonlocal signal emerges precisely at site $j$, its amplitude systematically decreases as $j$ approaches the center of the chain. To quantify this behavior, the lower-left panel displays the integrated signal intensity within the nonlocal cone -- summed over the bulk sites 10 to 90 to avoid edge effects -- as a function of the target position $j$ for various $\alpha$ values. As shown, the signal decays as a power law toward the middle of the chain before symmetrically increasing as we approach the opposite boundary. By fitting this algebraic decay (shown in the bottom-right panel), we extract the power-law exponent $p(\alpha) \approx 3 + 2\alpha$, in agreement with Eq.~\eqref{Intraband_scaling_SI}.  Deviations from the predictions for $j>N/2$ are associated with boundary conditions. See also Fig.~\ref{Prjb3} in the main text. \\

\begin{figure}[htbp!]
    \centering
    \includegraphics[width=\linewidth]{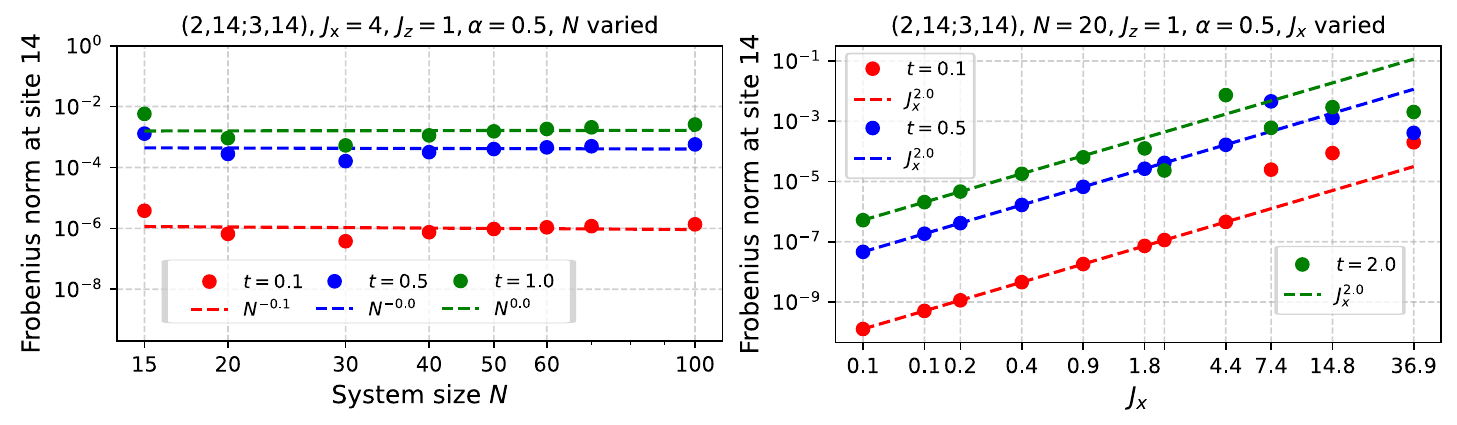}
    \caption{Frobenius norm at site 14 for $\alpha=0.5$ computed at different times, see legend, under the projected Hamiltonian ${\hat H}_{\mathrm{eff}}^{b=2}$. \textbf{Left:} dependence on system size $N$ for initial states in band 2: $(2,14;\,3,14)$ (notation defined in Eq.~\eqref{pairnotation} in the main text). The intraband nonlocal signal is independent of $N$,  as analytically expected from Eq.~\eqref{Intraband_scaling_SI}. 
    \textbf{Right:} dependence on $J_x$ at fixed $N=20$ for the same initial states, showing the predicted $\sim J_x^2$ scaling. Note that for large $J_x$, strong confinement as shown in Fig.~\ref{confinementLC} affects the strength of the light cone.} 
    \label{WintraScalingNoKac}
\end{figure}

\textbf{Dependence on $\boldmath{J_x}$, $\boldmath{N}$:} The scaling of this nonlocal signal is then studied as a function of system size \(N\) and coupling \(J_x\), as shown in Fig.~\ref{WintraScalingNoKac}. The left panel shows that the nonlocal signal is independent of \(N\). The right panel shows a clear \(\sim J_x^2\) scaling at fixed \(N\), consistent with Eq.~\eqref{Intraband_scaling_SI}. Deviations from the predictions at large \(J_x\) are associated with confinement effects (see Sec.~\ref{confinement}) arising from open boundary conditions.\\

\begin{figure}[htbp!]
    \centering
    \includegraphics[width=0.7\linewidth]{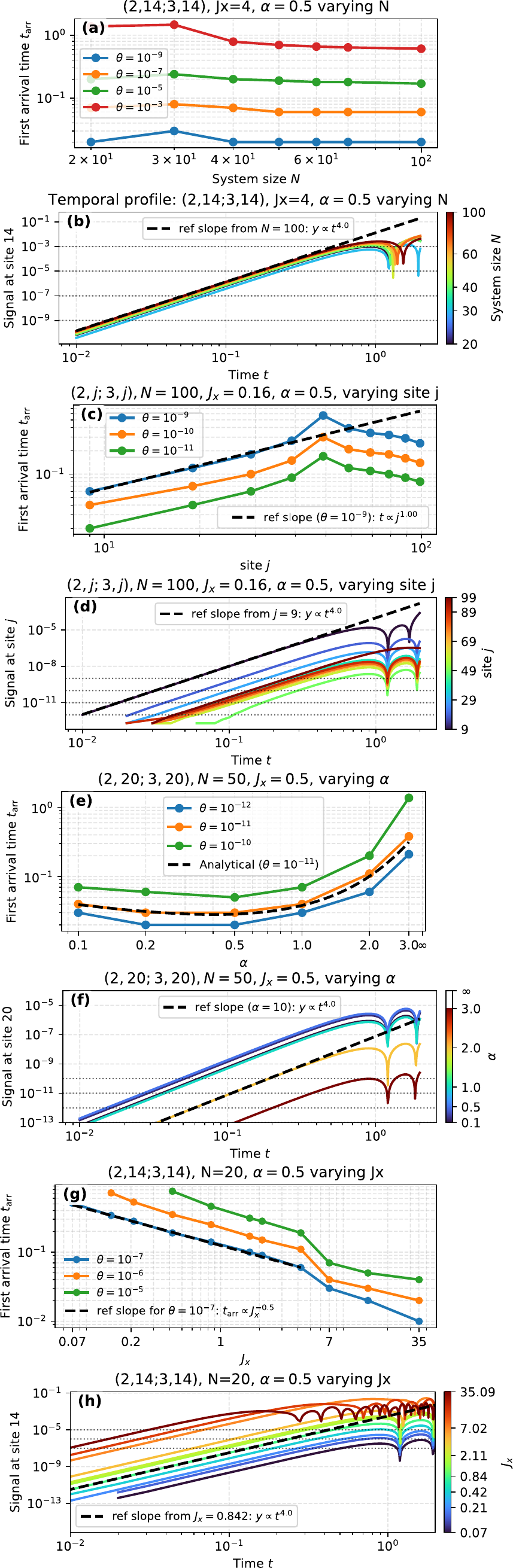}
    \caption{
    First-arrival times and temporal growth of the Frobenius norm signal at the nonlocal site $j$ for projected dynamics ${\hat H}_{\mathrm{eff}}^{b=2}$ for initial state in band 2: $(2,j;\,3,j)$ (notation defined in Eq.~\eqref{pairnotation} in the main text). Dashed lines denote reference fits. \textbf{(a)} First-arrival time $t_{\mathrm{arr}}$ at site $j=14$ as a function of system size $N$ for several detection thresholds $\theta$, at fixed $J_x=4$ and $\alpha=0.5$.  \textbf{(b)}  Corresponding temporal profiles $\|\Delta \rho_{14}(t)\|_{F}$.  \textbf{(c)} First-arrival time versus site index $j$ for fixed $N=100$, $J_x=0.16$, and $\alpha=0.5$, shown for multiple detection thresholds $\theta$. \textbf{ (d)} Corresponding temporal profiles $\|\Delta \rho_{j}(t)\|_{F}$ for varying site index $j$.   \textbf{(e)} First-arrival time at site $j=14$ as a function of interaction exponent $\alpha$ for fixed $N=50$ and $J_x=0.5$, shown for several detection thresholds $\theta$. Dashed line corresponds to analytics from Eq.~\eqref{tarr}.  \textbf{(f)}  Corresponding temporal profiles $\|\Delta \rho_{14}(t)\|_{F}$ for varying $\alpha$.  \textbf{(g)} First-arrival time at site $j=14$ as a function of interaction strength $J_x$ for fixed $\alpha$ and $N$, shown for several detection thresholds $\theta$.  \textbf{(h)}  Corresponding temporal profiles $\|\Delta \rho_{14}(t)\|_{F}$ colored by $J_x$.}
    \label{NonlocalarrivalTimes}
\end{figure}

\textbf{Dependence on time $t$:} 
To quantify the onset of the nonlocal signal, we define the first-arrival (or first-detection) time at the target site $j$, $t_{\mathrm{arr}}(j)$, as the moment the signal strength first crosses a chosen detection threshold $\theta$. Explicitly, it is given by the minimum time satisfying $\|\Delta \rho_j(t)\|_F \geq \theta$. Because the signal exhibits a rapid short-time power-law growth scaling as $t^4$, we vary $\theta$ over several orders of magnitude (typically from $10^{-13}$ to $10^{-3}$). This broad range ensures that our observations of the arrival dynamics are robust and independent of any specific choice of detector sensitivity.

The resulting first-arrival times are plotted in Figs.~\ref{NonlocalarrivalTimes}(a,c,e,g) as functions of the system size $N$, site index $j$, long-range exponent $\alpha$, and interaction strength $J_x$, respectively, for the various thresholds $\theta$ indicated in the legend. These numerical results show excellent agreement with the analytical predictions derived in Eq.~\eqref{Intraband_scaling_SI}. 

To complement these metrics, Figs.~\ref{NonlocalarrivalTimes}(b,d,f,h) display the explicit temporal evolution of the signal at the designated sites. The physical parameters and initial conditions in each of the panels (b,d,f,h) are identical to those in the arrival-time panel immediately preceding. In all cases, the short-time profiles confirm the $t^4$ scaling.
\\




\begin{figure}[htbp!]
    \centering
    \includegraphics[width=\linewidth]{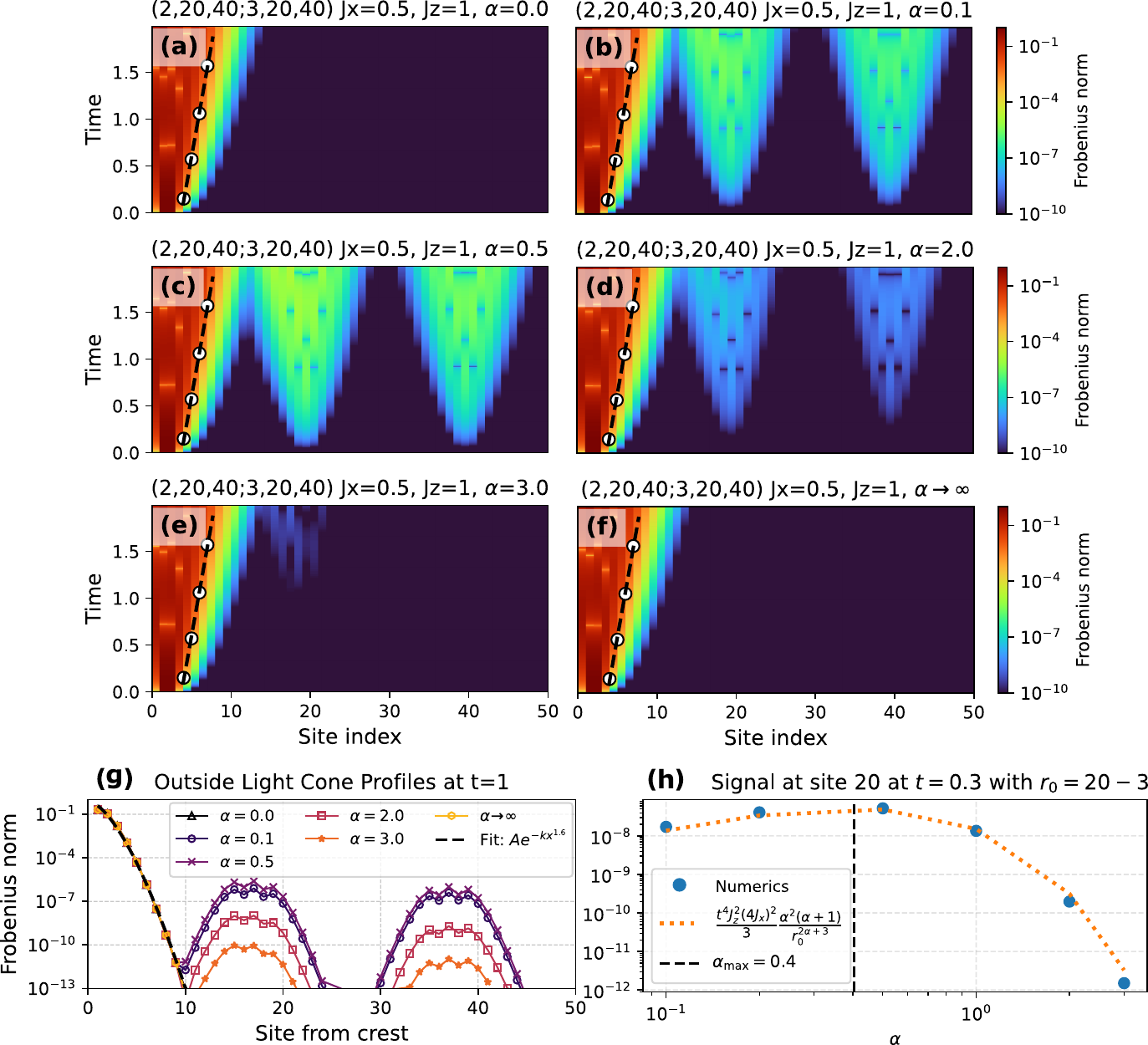}
    \caption{\textbf{(a-f)} Density plots of the Frobenius norm showing light-cone dynamics under the projected Hamiltonian ${\hat H}_{\mathrm{eff}}^{b=3}$ with initial states prepared in band~3 (2,20,40;\,3,20,40)  (notation defined in Eq.~\eqref{pairnotation} in the main text) with $ J_x=0.5$ and $N=50$. The six panels differ only in the value of $\alpha$. For $\alpha$=0 and $\alpha\to\infty$ (implemented by keeping only nearest-neighbor couplings), the nonlocal signal disappears. Panel \textbf{(g)} shows the light-cone profiles at $t=1$, where the secondary lobes peak at $\alpha \lesssim 1$ and weaken for $\alpha > 1$, while \textbf{(h)}  quantifies this by plotting the signal at site $j=20$ versus $\alpha$ at $t=0.3$. The dashed curve shows the analytic short-time prediction obtained from Eq.~\eqref{Intraband_scaling_SI}, where $r_0=|20-3|$ is the message-target separation. The black dashed vertical line corresponds to the analytical maximum from Eq.~\eqref{alphamax}.}
    \label{B0LFIMb2lcfarb3Varyalpha}
\end{figure}

\textbf{Fixed-time scaling with $\alpha$ and optimal interaction range $\alpha_{\mathrm{max}}$:} We now examine the dependence of the nonlocal signal on the long-range exponent \(\alpha\). From panels (a–f) in Figure~\ref{B0LFIMb2lcfarb3Varyalpha}, it is visually clear that the nonlocal signal is suppressed both in the fully connected limit ($\alpha=0$) and the short-range limit ($\alpha\to\infty$). The bottom-left panel (g) provides a spatial snapshot at $t=1$, where the nonlocal signal peaks at $\alpha \sim 0.5$. This is further quantified in the bottom-right panel (h), which tracks the signal at a specific target distant site $j=20$ together with the analytical prediction (dotted curve). As one can see, the analytical prediction from Eq.~\eqref{Intraband_scaling_SI} agrees very well with the numerical data. Moreover, the optimal interaction exponent predicted by Eq.~\eqref{alphamaxexact} is in agreement with the numerical results, see vertical dashed line in panel (h).

\subsection{Intraband signal under the Kac Prescription }
To analyze the thermodynamic scaling of the long-range interaction, we apply the standard Kac prescription  in the long-range case $0<\alpha < 1$~\cite{Botzung_2021,Mori_2012,kastner_2025},
\begin{equation}
\label{Kacresc}
J_x=\frac{J_{\mathrm{long}}}{\mathcal{N}_{\mathrm{Kac}}(\alpha,N)},
\qquad
\mathcal{N}_{\mathrm{Kac}}(\alpha,N) \equiv \mathcal{N}_\alpha
=
\frac{1}{N}\sum_{i<j}\frac{1}{|i-j|^\alpha},
\end{equation}
which ensures that the energy per site remains finite in the thermodynamic limit when $J_{\mathrm{long}}$ is kept constant. For $0<\alpha < 1$, one has 
$\mathcal{N}_{\mathrm{Kac}} \sim N^{1-\alpha}$. 

By substituting the Kac-rescaled coupling $J_x = J_{\mathrm{long}}/\mathcal{N}_{\mathrm{Kac}}$ into Eq.~\eqref{Intraband_scaling_SI}, the intraband signal becomes:
\begin{equation}
\label{KacintrawithN}
\tilde{\mathcal{F}}_{\mathrm{intraband}}^{(4)} \approx \frac{t^4 J_z^2 (4J_{\mathrm{long}})^2}{3 \mathcal{N}_{\mathrm{Kac}}^2}  \frac{\alpha^2(\alpha+1)}{r_0^{2\alpha+3}}\,.
\end{equation}
Consequently, under Kac rescaling, the intraband signal 
vanishes as $1/N^{2-2\alpha}$
 in the thermodynamic limit.

To determine the modified optimal interaction range exponent $\alpha$ from Eq.~(\ref{KacintrawithN}), we define 
\begin{equation}
\tilde{g}(\alpha)
=
\frac{(\alpha^3+\alpha^2)}{r_0^{2\alpha}\mathcal{N}_\alpha^2},
\end{equation}
and impose the stationarity condition
\begin{equation}
\frac{d}{d\alpha}\ln \tilde{g}(\alpha)=0,
\end{equation}
which gives
\begin{equation}
\frac{d}{d\alpha}
\left[
\ln(\alpha^3+\alpha^2)
-2\alpha\ln r_0
-2\ln \mathcal{N}_\alpha
\right]=0.
\end{equation}

This yields the implicit equation for $\alpha_{\mathrm{max}}$:
\begin{equation}
\label{maxforAlphamax}
\frac{3\alpha+2}{\alpha(\alpha+1)}
-2\ln r_0
-2 \frac{d}{d\alpha} \ln \mathcal{N}_\alpha
=0 \,.
\end{equation}

The Kac normalization for open boundary conditions is
\begin{equation}
\mathcal{N}_\alpha
=
\frac{1}{N}\sum_{r=1}^{N-1}\frac{N-r}{r^\alpha}
=
\sum_{r=1}^{N-1}r^{-\alpha}
-\frac{1}{N}\sum_{r=1}^{N-1}r^{1-\alpha}.
\end{equation}
For $0\le \alpha<1$, the leading asymptotic behavior is
\begin{equation}
\label{Kacopen}
\mathcal{N}_\alpha
\approx
\frac{N^{1-\alpha}}{(1-\alpha)(2-\alpha)} \,,
\end{equation}
and consequently,
\begin{equation}
   \frac{d}{d\alpha} \ln \mathcal{N}_\alpha
=
-\ln N
+\frac{1}{1-\alpha}
+\frac{1}{2-\alpha}.
\end{equation}

Substituting into Eq.~\eqref{maxforAlphamax} gives
\begin{equation}
\frac{3\alpha+2}{\alpha(\alpha+1)}
-2\ln r_0
+2\ln N
-\frac{2}{1-\alpha}
-\frac{2}{2-\alpha}
=0.
\end{equation}

In the thermodynamic limit $N /r_0 \to \infty$, we have  $2\ln N -2 \ln r_0 \to \infty$ for any fixed $0\leq\alpha<1$. The equation can only be solved near $\alpha=1$ where the term $2/(1-\alpha)$ diverges. We obtain
\begin{equation}
\ln \frac{N}{r_0}  \approx \frac{1}{1-\alpha} \,.
\end{equation}
Hence, the asymptotic value of $\alpha_{\mathrm{max}}$ with Kac rescaling in the thermodynamic limit is:
\begin{equation}
\label{alphamax with Kac}
\alpha_{\mathrm{max}} \approx 1 - \frac{1}{\ln (N/r_0)} \to 1 \,.
\end{equation}

\subsection{Numerical Study of the Intraband Signal With Kac Rescaling}

\begin{figure}[htbp]
\centering
\includegraphics[width=\linewidth]{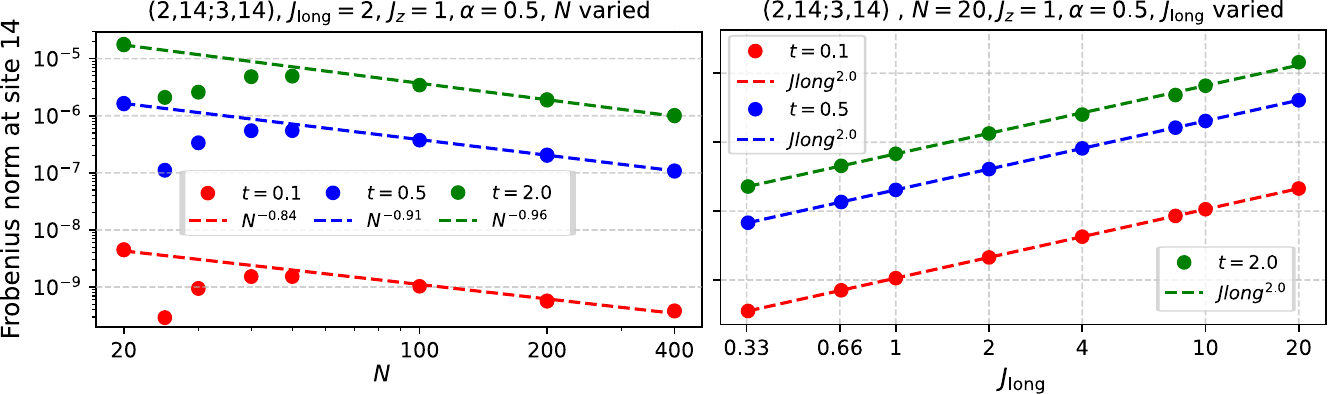}
\caption{Frobenius norm at site 14 and several different time slices $t$ for $\alpha=0.5$ under the projected Hamiltonian ${\hat H}_{\mathrm{eff}}^{b=2}$. \textbf{Left:} scaling with system size $N$, showing $\sim N^{-1}$ scaling for $N>100$. \textbf{Right:} scaling with $J_{\mathrm{long}}$ at fixed $N=20$, showing $\sim J_{\mathrm{long}}^2$ scaling. The behavior is consistent with Eq.~\eqref{KacintrawithN}.}
\label{WintraScaling}
\end{figure}

\textbf{Dependence on $N, J_x$:}
Figure~\ref{WintraScaling} quantifies the magnitude of the intraband signal for $\alpha=0.5$ with Kac rescaling under the projected Hamiltonian $\hat H_{\mathrm{eff}}^{b=2}$. With Kac rescaling, the signal decreases with system size as $\sim N^{-1}$ (left panel), consistent with Eq.~\eqref{KacintrawithN}, while at fixed system size it grows with interaction strength as $\sim J_{\mathrm{long}}^2$ (right panel).

\begin{figure}[htbp!]
\centering
\includegraphics[width=\linewidth]{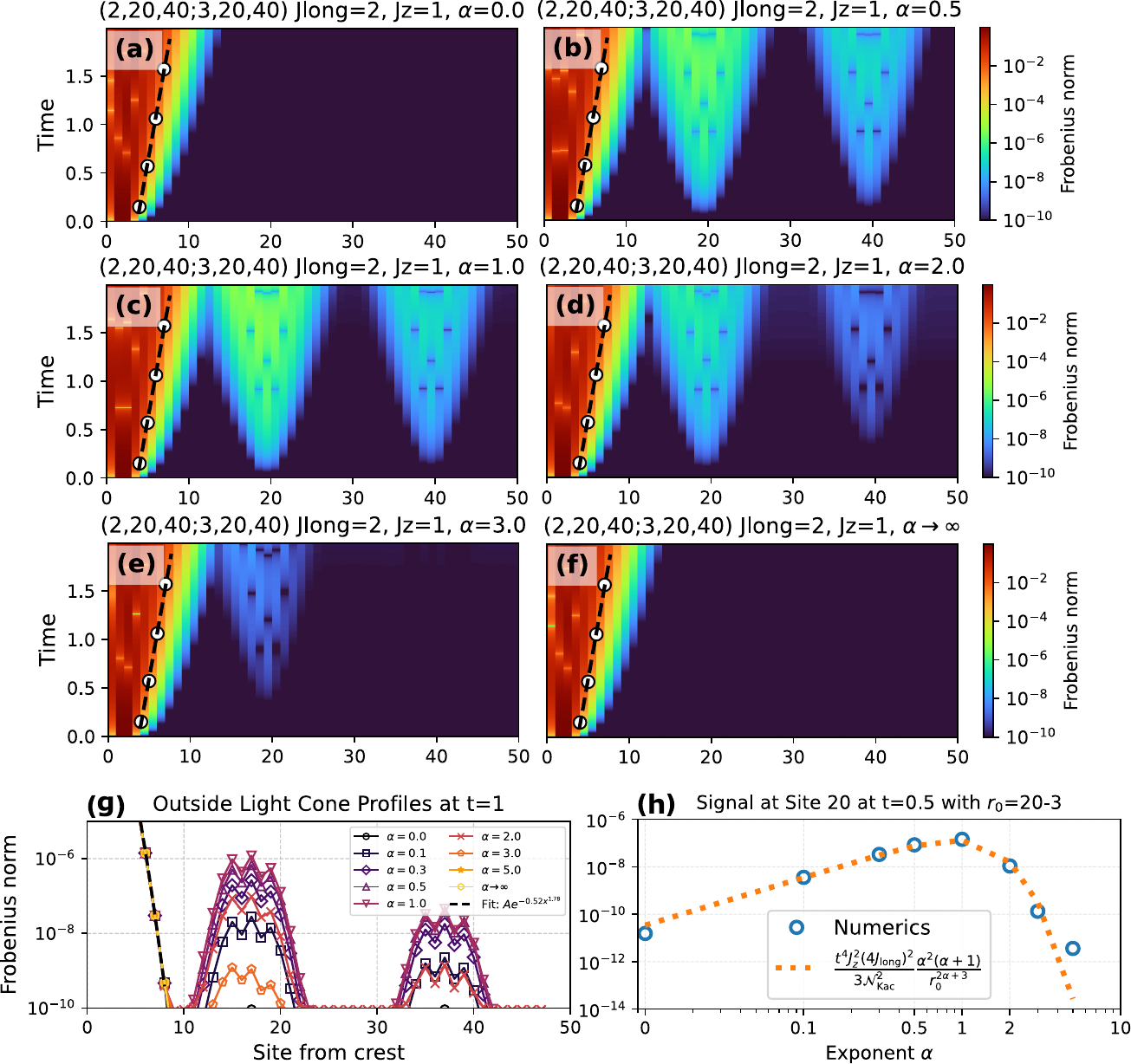}
\caption{Density plots of the Frobenius norm showing light-cone dynamics under the projected Hamiltonian ${\hat H}_{\mathrm{eff}}^{b=3}$ with initial states prepared in band~3 (2,20,40;\,3,20,40)  (notation defined in Eq.~\eqref{pairnotation} in the main text) with $ J_{\mathrm{long}}=2$ and $N=50$. The six panels \textbf{(a-f)} differ only in the value of $\alpha$. For $\alpha$=0 and $\alpha\to\infty$ (implemented by keeping only nearest-neighbor couplings), the nonlocal signal disappears. \textbf{(g)} shows the light-cone profiles at $t=1$, where the secondary lobes peak at $\alpha \sim 1$ and weaken for $\alpha > 1$ while \textbf{(h)} quantifies this by plotting the signal at site $j=20$ versus $\alpha$ at $t=0.5$. The dashed curve shows the analytic short-time prediction obtained from Eq.~\eqref{KacintrawithN}, where $r_0=|20-3|$ is the message-target separation.}
\label{B0LFIMb2lcfarb3VaryalphaJlong}
\end{figure}

\textbf{Dependence on $\alpha$: }The non-monotonic dependence on the interaction exponent $\alpha$ is analyzed in Fig.~\ref{B0LFIMb2lcfarb3VaryalphaJlong}. 
Here, we show the density plots of the Frobenius norm showing light-cone dynamics under the projected Hamiltonian for different $\alpha$ values, panels (a-f). One sees that for $\alpha=0, \infty$ (the latter implemented by keeping only nearest-neighbor couplings),
the nonlocal signal disappears. Moreover, in panel (g) we 
show the light-cone profiles at $t = 1$. Clearly, the nonlocal cones peak at $\alpha \sim 1$ and weaken in the short-range limit for $\alpha >1$. Lastly, in panel (h) we plot the signal at some specific site and time as a function of the interaction range $\alpha$. Results are shown together with our analytical predictions, see Eq.~\eqref{alphamax with Kac}.


\begin{figure}[htbp!]
\centering
\includegraphics[width=\linewidth]{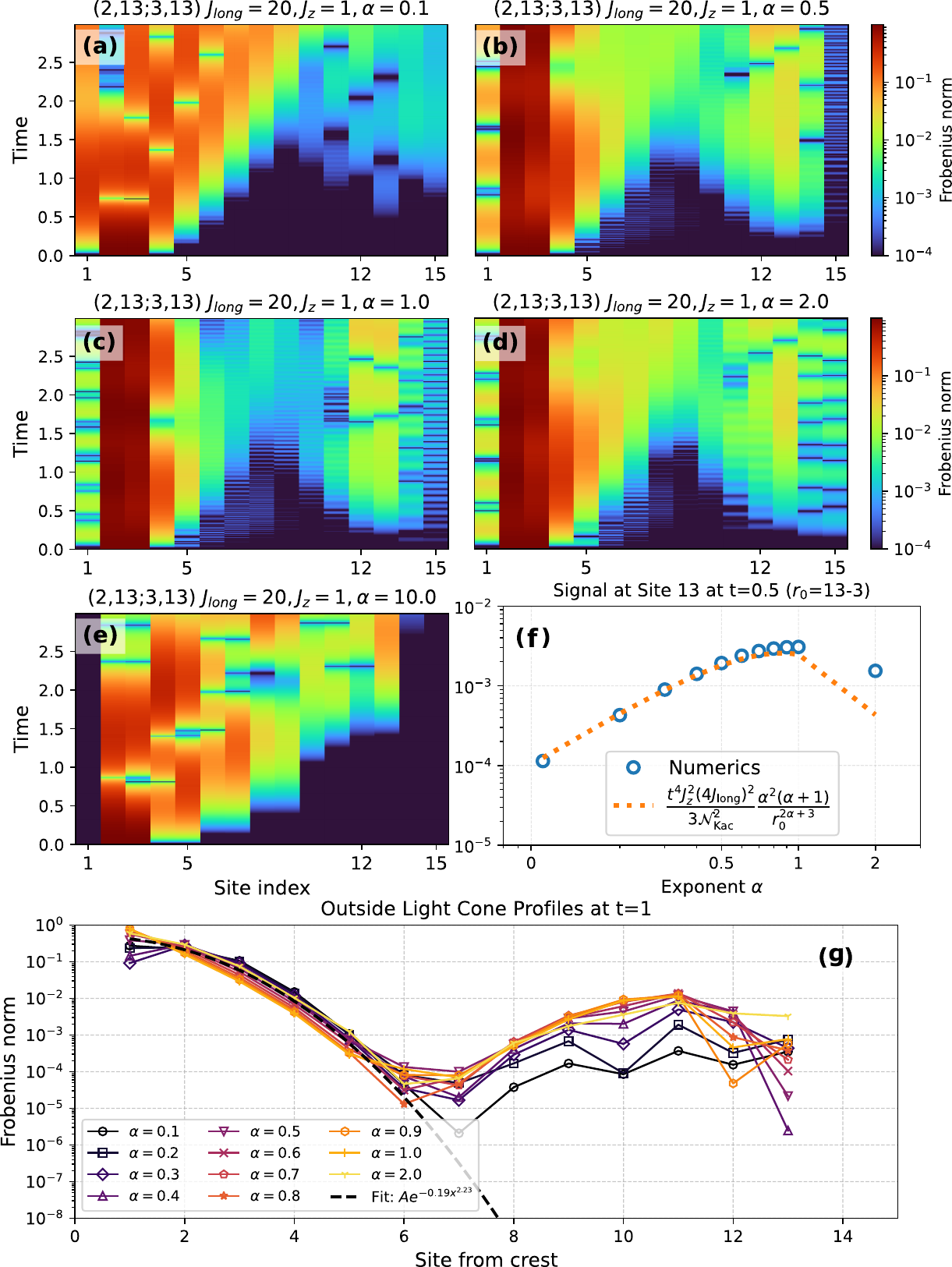}
\caption{\textbf{(a–e)} Density plots of the Frobenius norm showing light-cone dynamics under the full Hamiltonian for initial states prepared in band~2. \textbf{(f)} Frobenius-norm signal at site $j=13$ versus $\alpha$ at $t=0.5$. The dashed curve shows the analytic short-time prediction obtained from Eq.~\eqref{KacintrawithN}, where $r_0=|13-3|$ is the message-target separation. \textbf{(g)} Light-cone profiles at $t=1$ for all $\alpha$ values studied, illustrating the clear nonlocal signal, which should be experimentally observable. These profiles show that the signal outside the light cone decays superexponentially.}
\label{N15LFIMb2VaryalphaJlong20}
\end{figure}

Finally, in Fig.~\ref{N15LFIMb2VaryalphaJlong20} we verify that the results obtained 
in Fig.~\ref{B0LFIMb2lcfarb3VaryalphaJlong} using the band-projected Hamiltonian hold also for the full Hamiltonian. The density plots in panels (a–e) and the signal at site $j=13$ in panel (f) confirm that the maximal intraband signal is obtained around $\alpha \simeq 1$ in agreement with Eq.~(\ref{alphamax with Kac}).

\section{Derivation of the Interband Signal}
The short-time expansion accurately captures the intraband dynamics because the physics is confined to a fixed particle-number sector and only local virtual hopping processes are involved. However, such an expansion fails to correctly describe interband dynamics, which involve transitions between sectors with different particle numbers. The issue originates from the structure of the time-evolution operator, $U(t)=e^{-iHt}$. A finite-order Taylor expansion generates only polynomials in $H$ and therefore cannot produce the energy denominators associated with virtual transitions across large spectral gaps. Consequently, the short-time expansion cannot capture the strong suppression of interband processes caused by the large energy separation between particle-number manifolds. To correctly capture the interband contribution, one can instead use time-dependent perturbation theory (TDPT), where the suppression arises naturally through the energy gaps.

\subsection{The Hamiltonian}
We consider the full effective Hamiltonian of Eq.~\eqref{0thOrder} in Methods, which contains both intraband hopping and interband pair-creation processes. 
\begin{align}
{\hat H}^{(0)} ={}& 
J_z \sum_j \left( \hat{a}_j^\dagger \hat{a}_{j+1} + \hat{a}_{j+1}^\dagger \hat{a}_j + \hat{a}_j^\dagger \hat{a}_{j+1}^\dagger + \hat{a}_j \hat{a}_{j+1} \right) \nonumber \\
& -2 \sum_{i<j} \frac{J_x}{|i-j|^\alpha} (\hat{n}_i + \hat{n}_j) 
+ 4 \sum_{i<j} \frac{J_x}{|i-j|^\alpha} \hat{n}_i \hat{n}_j \nonumber \\
& + \sum_{i<j} \frac{J_x}{|i-j|^\alpha}.
\end{align}
We take the initial message state to be a single-particle excitation, $|i\rangle = |m\rangle.$ The pair-creation operator $a_{j}^\dagger a_{j+1}^\dagger$ couples the initial state $|i\rangle = |m\rangle$ to the three-particle state $|f\rangle = |m,r-1,r\rangle.$

\subsection{TDPT and the Interband Energy Gap}

Under first-order time-dependent perturbation theory (TDPT), the interband pair-creation process is modeled as a two-level transition driven by the weak off-diagonal coupling $J_z$. The time-dependent probability $P_m(t)$ of finding the newly created pair at the target sites $r-1$ and $r$ takes the classic form of a detuned Rabi oscillation,
\begin{equation}
\label{RabiFormula}
P_m(t) = \frac{4J_z^2}{\omega_m^2} \sin^2\left(\frac{\omega_m t}{2}\right) \,,
\end{equation}
where the Rabi frequency is governed by the energy detuning between the initial single-particle configuration and the final three-particle configuration, $\omega_m = E_f - E_i$. This leads to  
\begin{equation}
\label{Rabitimeavg}
A(\omega_m) = \frac{2J_z^2}{\omega^{2}_m}
\end{equation}
as the steady-state, time-averaged probability, which filters out the fast time-dependent Rabi oscillations.

To evaluate $\omega_m$ explicitly, we partition the diagonal parts of the long-range Hamiltonian $\hat{H}^{(0)}$ into three distinct physical contributions: the bare constant background lattice energy $E_{b=0}$, the single-body potential $V$, and the two-body density-density interactions $W$ (such that $E = E_{b=0} + V + W$).

For the initial state in band 1,  $|i\rangle = |m\rangle$:
\begin{align}
    E_{b=0} &= \frac{J_x}{2} \sum_{i \neq j} |i-j|^{-\alpha}\\
    V^{(i)} &= -2J_x \sum_{j \neq m} |m-j|^{-\alpha} = V_m\\
    W^{(i)} &= 0
\end{align}
Because there is only one particle in the system, no density-density pairs exist.\\

For the final state in band 3, $|f\rangle = |m,r-1,r\rangle$, the baseline energy $E_{b=0}$ is unchanged. The single-body potential is now the sum of the potentials at all three occupied sites: $V^{(f)} = V_m +  V_{r-1}+ V_r$. The density-density interaction term $W^{(f)}$ sums over the three unique interacting pairs formed by the particles: $W^{(f)} = W_{r,r-1} + W_{m,r} + W_{m,r-1}$.\\

Taking the exact energy difference $\omega_m = E_f - E_i$,
\begin{equation}
\label{exactfreq}
\omega_m =  V_r + V_{r-1} + W_{r,r-1} + W_{m,r} + W_{m,r-1}.
\end{equation}
Here, $W_{r,r-1} = 4J_x / 1^\alpha = 4J_x$ between the newly created adjacent excitations. Also, notice that $V_m$ disappears from the gap equation, which means that the position-dependent potential terms are identical for $m=3$ and $m=2$.

We therefore split Eq.~\eqref{exactfreq} into terms $E_0$ that do not depend on the message position $m$ and terms $\Delta W_m$ that do depend on  $m$:
\begin{align}
\omega_m &= E_0+\Delta W_m \,, \notag \\
\label{omegam} E_0 &= V_r + V_{r-1} + W_{r,r-1} \,,  \\
\Delta W_m &= W_{m,r} + W_{m,r-1} \,. \notag
\end{align}
Only $\Delta W_m$ depends on the position $m$ of the initial state excitation.

\subsection{Taylor Expansion and Validity Regime}
\label{taylorexpansion}

To isolate the signal difference between two distinct message locations (e.g., $m=2$ and $m=3$), we look at the change in the time-averaged magnetization at site $r$ associated with the initial excitation shifting from site $m=2$ to $m=3$: $2|A_{\omega_2}- A_{\omega_3} |$. The factor of 2 is due to the existence of another process where a pair is created at sites $r$, $r+1$ instead of $r-1$, $r$, which for large $r$ contributes with the same small probability. 

The time averaging $\langle \sin^2(\omega_m t / 2) \rangle \to 1/2$ leading from Eq.~\eqref{RabiFormula} to Eq.~\eqref{Rabitimeavg} is valid when the integration window spans many oscillation cycles: $T_{\text{avg}} \gg 2\pi / E_0$.
Furthermore, if we focus on intermediate times $2\pi / E_0 \ll T_{\text{avg}} \ll  2\pi /\Delta W_m$, the Rabi oscillations associated with frequencies $\omega_2$ and $\omega_3$ in Eq.~\eqref{RabiFormula} remain in phase with one another, and thus the time-averaged Frobenius norm (the time average of the absolute value of the magnetization difference) is equivalent to the absolute value of the time-averaged magnetization difference, 
\begin{equation}
\mathcal{F}_{\mathrm{interband}} = 2|A_{\omega_2}- A_{\omega_3} | \,.
\end{equation}
As we will see below, $E_0 \sim  J_x N^{1-\alpha}$ and  $\Delta W_m \sim \alpha J_x/r^{\alpha+1}$, so the explicit regime of validity of this approximation is $2\pi / (J_x N^{1-\alpha})\ll T_{\text{avg}} \ll 2\pi r^{\alpha+1} /(\alpha J_x)$.

Because we care about the dominant scaling with system size $N$, we can approximate the discrete sum over the lattice coordinates in each $V$ term as a continuous integral. Then $V_r$, $V_{r-1}$ in Eq.~\eqref{omegam} each scales as $J_x N^{1-\alpha}$, e.g., $V_r \approx V_{r-1} \approx -4 J_x N^{1-\alpha}/(1-\alpha)$ for periodic boundary conditions or for open boundary conditions when the site $r$ is far from the boundary.
On the other hand, the density-density terms $W$ are $O(J_x)$, not growing with $N$. Consequently, $E_0 \sim J_xN^{1-\alpha},$ and the position-dependent correction satisfies $ |\Delta W_m| \ll |E_0|$ in the perturbative regime. We perform a first-order Taylor expansion of the time-averaged probability function $A(\omega)$ around $ E_0$:
\begin{equation}
A(E_0 + \Delta W_m) \approx A(E_0) + \left. \frac{\partial A}{\partial \omega} \right|_{E_0} \Delta W_m.
\end{equation}

Evaluating the difference $\Delta A = A(\omega_3) - A(\omega_2)$, the $A(E_0)$ term cancels out. Differentiating the probability function yields $\partial A / \partial \omega = -4J_z^2 \omega^{-3}$, which gives:
\begin{equation}
|\Delta A| \approx |\Delta W_3 - \Delta W_2| \frac{4J_z^2}{|E_0|^3}.
\end{equation}
We note that boundary conditions enter here only through the single-body potential terms $V_r + V_{r-1}$ inside the denominator $E_0$; the boundary conditions do not affect the scaling behavior.

At large physical separations $r$ between the message coordinate and the pair-creation site, the discrete difference between the local interactions can be approximated by the first derivative, 
\begin{equation}
    |\Delta W_3 - \Delta W_2| \approx  \frac{\partial W}{\partial m} \approx \frac{4 \alpha J_x}{r^{\alpha+1}} \,.
\end{equation}
For periodic boundary conditions, or for $r$ far from the boundary in the case of open boundary conditions, $E_0 \approx V_r +V_{r-1} \approx 8 J_x N^{1-\alpha}/(1-\alpha)$, and thus
\begin{equation}
\label{Interband_SI}
\mathcal{F}_{\mathrm{interband}} \approx \frac{(1-\alpha)^3}{16} \frac{J_z^2 \left( \alpha J_x / r^{\alpha+1} \right)}{\left( J_x N^{1-\alpha} \right)^3} \propto \frac{ J_z^2}{J_x^2 N^{3-3\alpha} r^{\alpha+1}} \,.
\end{equation}

Although this scaling was derived explicitly for an initial state in band 1, the argument extends directly to any band. For example, when initializing in band 2, Eq.~(\ref{omegam}) contains additional contributions, but each of them scales as $J_x N^{1-\alpha}$. Consequently, the system-size dependence remains unchanged; only the overall prefactor, and therefore the signal magnitude, is modified.

\subsection{Modifications Under the Kac Prescription}
The Kac prescription 
(in the regime $0<\alpha < 1$) modifies the system by rescaling the long-range interaction strength, see Eq.~\eqref{Kacresc}, where the Kac factor $\mathcal{N}_\alpha$ scales as $N^{1-\alpha}$ for large $N$, and is, for example, given by Eq.~\eqref{Kacopen} for open boundary conditions. This sensibly changes the scaling of the interband signal because the macroscopic energy gap is regularized. With the substitution of Eqs.~\eqref{Kacresc} and \eqref{Kacopen}, Eq.~\eqref{Interband_SI} becomes
\begin{equation}
\label{Interband_scaling_Kac_SI}
\tilde{\mathcal{F}}_{\mathrm{interband}} \approx \frac{\alpha }{16(2-\alpha)^3} \frac{J_z^2}{ J_{\mathrm{long}}^2 N^{1-\alpha} r^{\alpha+1}}  \propto \frac{ J_z^2}{J_{\mathrm{long}}^2 N^{1-\alpha} r^{\alpha+1}} \,.
\end{equation}
For periodic boundary conditions, the prefactor is modified, but the scaling behavior is unchanged.
In effect, the Kac prescription replaces the  $1/N^{3-3\alpha}$ scaling in Eq.~\eqref{Interband_SI} caused by the divergent energy gap, replacing it with a much milder $1/N^{1-\alpha}$ scaling arising purely from the weakened interaction strength. While the interband leakage still strictly vanishes in the thermodynamic limit, the Kac prescription causes it to vanish at a significantly slower rate.

\subsection{Numerical Investigation of the Interband Signal}
\label{interband}
Here we test our analytical results for the interband signal numerically, specifically checking the spatial dependence and the dependence on $J_x$, $N$, and $\alpha$. 

\begin{figure}[htbp!]
    \includegraphics[width=0.7\linewidth]{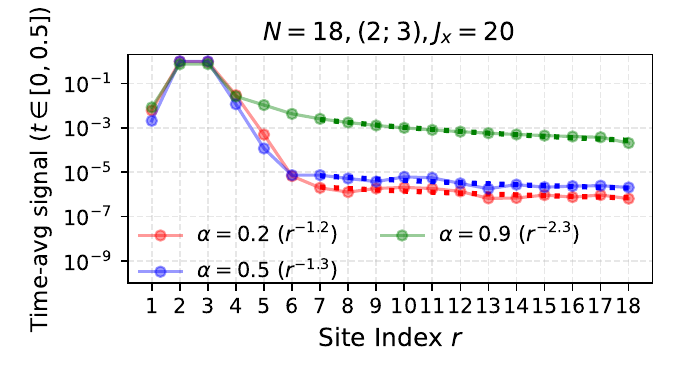}
    \includegraphics[width=0.7\linewidth]{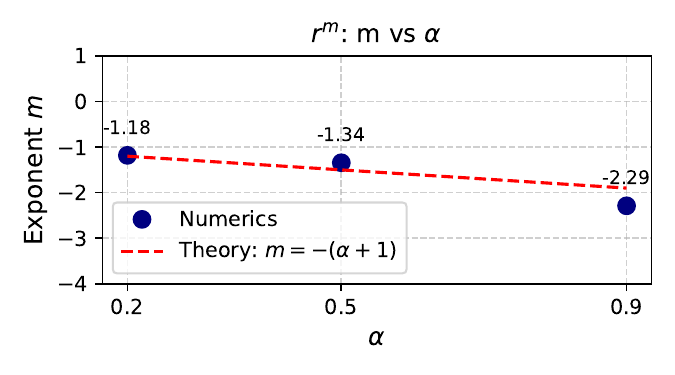}
    \caption{\textbf{Top:} Time-averaged interband Frobenius norm in the interval ${t\in[0,0.2]},$ as a function of distance $r$ for open boundary conditions with $N=18$, $J_x=20$, and initial states $(2;3)$ in band~1. Curves are shown for $\alpha=0.2,0.5,$ and $0.9$. Dotted lines denote power-law fits over the intermediate-distance fitting window used in the analysis. The extracted exponents are consistent with an approximate decay $\sim r^{-(\alpha+1)}$, as given by Eq.~\eqref{Interband_SI}. \textbf{Bottom: } Extracted spatial decay exponent $m$ obtained from power-law fits, for the data shown in the panel on top. Numerical fit exponents are plotted as a function of $\alpha$ and compared against the analytical prediction $m=-\alpha-1$ (dashed line). The agreement supports the TDPT prediction that for $\alpha >0$, the interband signal falls off with distance as $r^{-(\alpha+1)}$.}
    \label{InterbandSpatialDecay}
\end{figure}

\textbf{Scaling with distance:} We compute the time-averaged Frobenius norm, ${t\in[0,0.2]},$ for  open boundary conditions (OBC) at fixed interaction strength $J_x=5$ and system size $N=18$. The upper panel of Fig.~\ref{InterbandSpatialDecay} shows the spatial dependence of the signal for several long-range interaction exponents $\alpha$. The signal decreases with distance from the initial excitation and is well described by an approximate power-law decay over intermediate distances. Fitting the tails of the numerical data with a power law   $\sim r^m$ yields exponents that for $0<\alpha<1$ closely follow the TDPT prediction of Eq.~\eqref{Interband_SI}
with $m=-\alpha-1$. This behavior is summarized in the lower panel of Fig.~\ref{InterbandSpatialDecay}, where the numerically extracted exponents are compared directly against the analytical prediction. \\

\begin{figure}[htbp!]
    \includegraphics[width=0.7\linewidth]{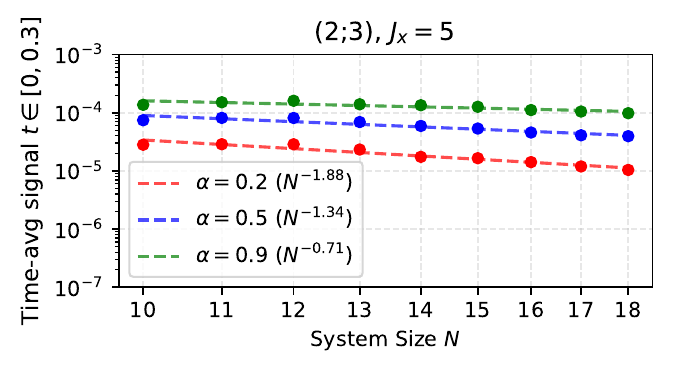}
    \includegraphics[width=0.7\linewidth]{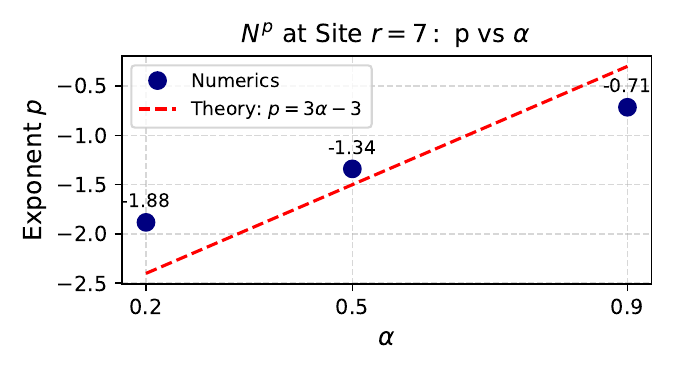}
    \caption{\textbf{Top:} Time-averaged interband signal, ${t\in[0,0.2]}$, at fixed probe site $r=7$ as a function of system size $N$ for periodic boundary conditions with initial states $(2;3)$ and $J_x=5$. Curves are shown for $\alpha=0.2,0.5,$ and $0.9$. Dashed lines denote power-law fits of the form $\sim N^p$. \textbf{Bottom: } Extracted system-size scaling exponent $p$ obtained from power-law fits, $N^p$, for the data shown in the top panel. Numerical exponents are plotted as a function of $\alpha$ and compared against the TDPT prediction $p=3\alpha-3$ (dashed line). The observed agreement supports the analytical prediction.}
    \label{WinterNExponent}
\end{figure}

\textbf{Scaling with $N$:}
To study the thermodynamic suppression of interband nonlocality, we compute the time-averaged Frobenius norm, ${t\in[0,0.2]}$, at a fixed probe site $r=7$ while varying the system size $N$ under periodic boundary conditions for the initial states $(2;3)$ in band 1. The upper panel of Fig.~\ref{WinterNExponent} shows that the interband signal decreases systematically with increasing system size for all interaction exponents studied. Power-law fits of the form $\sim N^p$ support Eq.~\eqref{Interband_SI}. The extracted scaling exponents are summarized in the lower panel of Fig.~\ref{WinterNExponent}, where they are compared against the analytical TDPT prediction of Eq.~\eqref{Interband_SI}, $p=3\alpha-3$. 
The numerical results are broadly consistent with the predicted scaling relation.\\

\textbf{Scaling with $\alpha$:}
In the strongly long-range regime ($\alpha \ll 1$), the  $\alpha$ dependence in the denominator of Eq.~\eqref{Interband_SI}, $N^{3-3\alpha}r^{\alpha+1}$, is subleading compared to the linear $\alpha$ factor in the numerator. The dominant overall scaling behavior is then governed strictly by $\ln \mathcal{F}_{\mathrm{interband}} \sim \ln \alpha$ or $\mathcal{F}_{\mathrm{interband}} \sim \alpha$. This scaling behavior is corroborated numerically in Fig.~\ref{N20LFIMb2VaryalphaJx5}(h) in the main text, where the time-averaged interband signal at site $j=10$ is plotted against $\alpha$, demonstrating that the signal strength scales linearly as $\sim \alpha$ for $\alpha < 1$. For $\alpha > 1$, the factors in the denominator of Eq.~\eqref{Interband_SI} dominate the $\alpha$ dependence, and the signal rapidly decays to zero.

\begin{figure}[h!]
    \centering
    \includegraphics[width=0.8\linewidth]{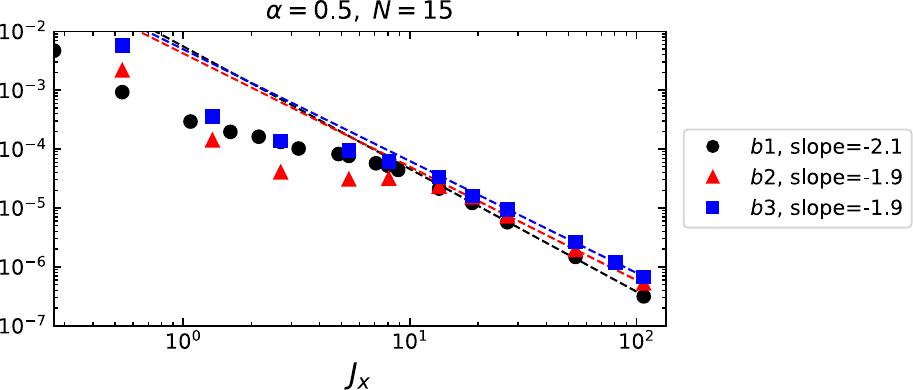}
    \caption{Interband nonlocality at $\alpha=0.5$ scales as $\|\Delta \rho_n(t)\|_F^{\mathrm{interband}} \sim J_x^{-2}$.
    The signal at site~13, time-averaged over ${t\in[0,2]}$, is shown for configurations initialized to $b_{1}(2;3)$, $b_{2}(2,5;3,5)$ and $b_{3}(2,5,7;3,5,7)$ for $N=15$.}
    \label{WinterScaling2}
\end{figure}

\textbf{Scaling with $J_x$:}
Fig.~\ref{WinterScaling2} shows the Frobenius norm signal at a fixed site $r=13$ for $N=15$ and $\alpha=0.5$. In addition to the band-1 initial states, results are shown for initial states in bands 2 and 3. In all cases, the initial excitations are far from site $13$, so intraband contributions are negligible, and the interband signal dominates. In all cases  we observe $\sim J_x^{-2}$ scaling of the signal for large $J_x$, consistent with Eq.~\eqref{Interband_SI}.

\subsection{The Interband Signal for $\alpha=0$}
\label{interbandalpha0}

\begin{figure}[h!]
    \centering
    \includegraphics[width=0.8\linewidth]{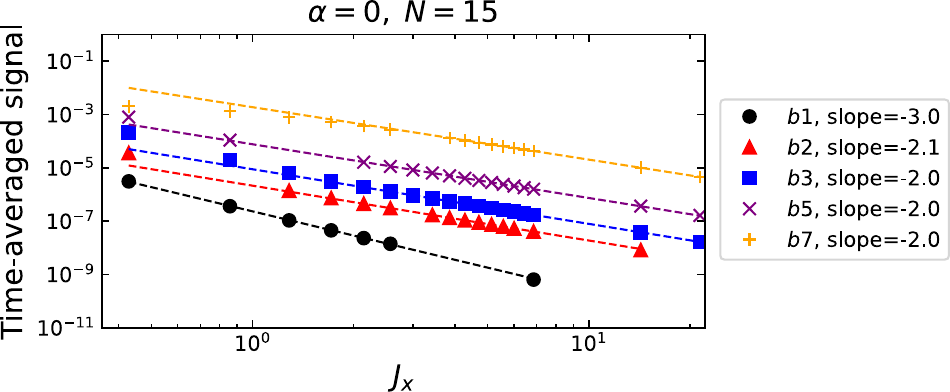}
    \caption{Interband nonlocality at $\alpha=0$ scales as $\|\Delta \rho_n(t)\|_F^{\mathrm{interband}} \sim J_x^{-2}$.
    The signal at site~13, time-averaged over ${t\in[0,2]}$, is shown for configurations initialized to $b_{1}(2;3)$, $b_{2}(2,5;3,5)$, $b_{3}(2,5,7;3,5,7)$, $b_{5}(2,5,7,9,11;3,5,7,9,11)$, and $b_{7}(2,5,7,9,11,13,15;3,5,7,9,11,13,15)$ for $N=15$.}
    \label{WinterScaling}
\end{figure}

The interband signal in 
Eq.~\eqref{Interband_SI} vanishes for the special case of $\alpha=0$. Physically, this is due to the fact that the two-body interaction is all-to-all, independent of distance, and thus $\Delta W_m$ in Eq.~\eqref{omegam} does not actually depend on the initial excitation position $m$. Thus, one may think that the interband signal vanishes in this special case. However, in Fig.~\ref{WinterScaling}, we see that there is indeed a nonzero signal, and the signal decays either with the same $J_x^{-2}$ scaling as in the $\alpha>0$ case (illustrated in Fig.~\ref{WinterScaling2}) or with the faster $J_x^{-3}$ decay for $b=1$. Despite an apparent similarity, the origin of the interband signal is quite different for $\alpha=0$. 

The dominant effect for $\alpha>0$, considered in Sec.~\ref{taylorexpansion} above, arises from jumps out of the initial band, e.g., from band 1 to band 3, through creation of an excitation pair. For $\alpha=0$, on the other hand, the dominant process is a virtual process where a pair is destroyed and another pair is created, ending up in a final state that is in the same band as the initial state (with no energy gap).

As a toy model, one can take periodic boundary conditions (so that the $V$ term in the Hamiltonian is position-independent) and consider a 4-dimensional Hilbert space spanned by the vacuum state $|b=0\rangle$ and three band-2 states that are separated from it by energy $4 N J_x$: $|1,2\rangle$, $|1,3\rangle$, and $|r-1,r\rangle$ for some $r\ge 5$. States $|1,2\rangle$ and $|r-1,r\rangle$ are directly coupled to the vacuum state $|b=0\rangle$ via pair creation/annihilation, proportional to $J_z$, while states $|1,2\rangle$ and $|1,3\rangle$ are coupled to each other by the hopping term, also $\propto J_z$. Here we start in state $|1,2\rangle$ or $|1,3\rangle$ and study the probability of an excitation appearing at site $r$, which in the toy model is just the probability of being in state $|r-1,r\rangle$. In this case, one can confirm analytically that after averaging over oscillations at short times $t \sim 1/(N J_x)$, the time-averaged probabilities and their difference scale as $J_z^2/(N J_x)^2$.

The scaling $\mathcal{F}_{\mathrm{interband}} \sim J_x^{-2}$ for $\alpha=0$ is observed in Fig.~\ref{WinterScaling}, for initial states in bands 2, 3, 5, and 7. In each case, the interband signal arises from a virtual process where an excitation pair in an initial state in band $b$ is annihilated and a distant pair at sites $r-1,r$ is created. Notably, the distant pair can be created at any distance with equal probability; thus, the interband signal for $\alpha=0$ is position-independent, in contrast with the $\sim r^{-1-\alpha}$ falloff with distance from the source in the $\alpha>0$ case, seen in Eq.~\eqref{Interband_SI}.

Of course, this virtual process is not available for initial states in band $b=1$, since there is no initial pair to annihilate. In this special case, the $\sim J_x^{-2}$ interband signal is absent, and the leading contribution to the Frobenius norm signal appears at higher order, $\mathcal{F}_{\mathrm{interband}} \sim J_x^{-3}$.

\section{Competition Between Intraband and Interband Dynamics and the Critical Controllability Regime}
\label{sec:competition_intraband_interband}

In this section, we estimate  the parameter space wherein the causal landscape can be controllable, meaning that the programmable nonlocal signal dominates over the unprogrammable background noise. To estimate a boundary, we compare the algebraic decay profiles of the intraband and interband Frobenius norms derived in the preceding sections. 

A key conceptual distinction between the two behaviors lies in their respective temporal evolution. The unprogrammable interband signal $F_{\mathrm{interband}}$ is driven by virtual transitions across a large energy gap $\Delta E \sim J_x N^{1-\alpha}$. It undergoes extremely fast fluctuations on a microscopic timescale $\tau \sim 1/\Delta E$, quickly saturating to a persistent, time-averaged steady-state background floor. Conversely, the programmable intraband signal $F_{\mathrm{intraband}}$ represents resonant, physical particle transport governed by the kinetic hopping rate $J_z$. It grows continuously for short times as a power law ($t^4$), and we can evaluate it at the relevant transport timescale $t \sim 1/J_z$.

Therefore, a controllable causal light cone requires that at the characteristic transport timescale, the programmable signal must overcome the time-averaged interband noise floor:
\begin{equation}
F_{\mathrm{intraband}}\left(t \sim \frac{1}{J_z}\right) > F_{\mathrm{interband}}.
\end{equation}
Dropping purely numerical prefactors to focus on the scaling with respect to the system parameters, the two foundational behaviors follow:
\begin{align}
F_{\mathrm{intraband}}(t) &\sim t^4 J_z^2 J_x^2 r_0^{-(2\alpha+3)}, \\
F_{\mathrm{interband}} &\sim J_z^2 J_x^{-2} N^{-3(1-\alpha)} r_0^{-(\alpha+1)},
\end{align}
where $\alpha < 1$ represents the long-range interaction exponent, $N$ is the system size, and $r_0$ is the spatial separation between the source and target sites. 

Below, we estimate an upper bound for the critical radius below which the causal landscape is controllable without and with the Kac rescaling prescription. 

\subsection{Regime Without Kac Rescaling}

When the long-range coupling strength $J_x$ is held constant as $N \to \infty$, substituting the characteristic transport time $t = 1/J_z$ into the short-time expansion of the intraband signal yields:
\begin{equation}
F_{\mathrm{intraband}}\left(t \sim \frac{1}{J_z}\right) \sim J_z^{-2} J_x^2 r_0^{-(2\alpha+3)}.
\end{equation}
Demanding that this signal dominates the interband background gives the inequality:
\begin{equation}
J_z^{-2} J_x^2 r_0^{-(2\alpha+3)} > J_z^2 J_x^{-2} N^{-3(1-\alpha)} r_0^{-(\alpha+1)}.
\end{equation}
Isolating the distance variable $r_0$ allows us to extract a critical distance, $r_c$, below which the system remains controllable:
\begin{equation}
r_0^{\alpha+2} < \left(\frac{J_x}{J_z}\right)^4 N^{3(1-\alpha)},
\end{equation}
which provides the scaling behavior for the critical radius of controllability:
\begin{equation}
r_c \sim \left[ \left(\frac{J_x}{J_z}\right)^4 N^{3(1-\alpha)} \right]^{\frac{1}{\alpha+2}}.
\end{equation}
Because $1-\alpha > 0$ for all long-range profiles in this regime, the term $N^{3(1-\alpha)}$ diverges in the thermodynamic limit. As a result, $r_c \to \infty$ as $N \to \infty$. This implies that without Kac rescaling, the unprogrammable background noise is perfectly suppressed by the interband energy gap, rendering the entire macroscopic length of the chain controllable. 

\subsection{Regime with Kac Rescaling}

To maintain an extensive total energy as the system size scales up, the standard Kac prescription rescales the long-range coupling parameter such that $J_x \sim J_{\mathrm{long}} / N^{1-\alpha}$, where $J_{\mathrm{long}}$ is kept constant. Under this transformation, the scaling profiles of our signals are altered:
\begin{align}
F_{\mathrm{intraband}}\left(t \sim \frac{1}{J_z}\right) &\sim J_z^{-2} J_{\mathrm{long}}^2 N^{-2(1-\alpha)} r_0^{-(2\alpha+3)}, \\
F_{\mathrm{interband}} &\sim J_z^2 J_{\mathrm{long}}^{-2} N^{-(1-\alpha)} r_0^{-(\alpha+1)}.
\end{align}
Notice that due to the regularization of the global energy gap, the interband noise now falls off much more slowly with system size ($\sim N^{-(1-\alpha)}$). Crucially, the programmable intraband signal now explicitly decays with system size as $N^{-2(1-\alpha)}$. 

Setting up the inequality under the Kac prescription gives:
\begin{equation}
N^{-2(1-\alpha)} r_0^{-(2\alpha+3)} > \left(\frac{J_z}{J_{\mathrm{long}}}\right)^4 N^{-(1-\alpha)} r_0^{-(\alpha+1)}.
\end{equation}
Solving for the new boundary of the controllable region gives:
\begin{equation}
r_0^{\alpha+2} < \left(\frac{J_{\mathrm{long}}}{J_z}\right)^4 \frac{1}{N^{1-\alpha}},
\end{equation}
so the critical distance under Kac rescaling becomes:
\begin{equation}
r_c \sim \left[ \left(\frac{J_{\mathrm{long}}}{J_z}\right)^4 \frac{1}{N^{1-\alpha}} \right]^{\frac{1}{\alpha+2}}.
\end{equation}

Since $1-\alpha > 0$, in the strict thermodynamic limit ($N \to \infty$), the critical radius shrinks to zero ($r_c \to 0$). This occurs because the programmable signal decays faster with system size than the unprogrammable background noise floor ($N^{-2(1-\alpha)} \ll N^{-(1-\alpha)}$). 

Thus, under a standard Kac rescaling, localized nonlocality cannot survive in the infinite-size limit for a fixed $\alpha < 1$. 

\section{Validity of the Projected Hamiltonian and Interband Leakage}

To justify the use of the projected Hamiltonian $\hat{H}^b_{\mathrm{eff}}$, we analyze the stability of the isolated band $b=1$ against leakage into the $b=3$ manifold.


Using first-order time-dependent perturbation theory, the transition probability from the initial state in band $b$ to a specific final state $|\psi_f\rangle$ in band  $b+2$ is given by the time-averaged rate:
\begin{equation}
    \overline{P}_{i \to f} \approx \frac{2 |\langle \psi_f | \hat{H} | \psi_i \rangle|^2}{\Delta E^2},
\end{equation}
where $\Delta E \sim J_x N^{1-\alpha}$ represents the energy gap between bands. Given the interband coupling strength $|\langle \psi_f | \hat{H} | \psi_i \rangle| \sim J_z$ and the presence of $\sim N$ reachable states in the higher manifold, the total global leakage probability $P_{\mathrm{leak}} = \sum_f \overline{P}_{i \to f}$ scales as:
\begin{equation}
    P_{\mathrm{leak}} \sim \left( \frac{J_z}{J_x} \right)^2 N^{2\alpha - 1}.
\end{equation}
For $\alpha > 1/2$, this global probability diverges with system size $N$, suggesting that the global wave function mixes significantly. 

Despite this global mixing, the projected Hamiltonian provides a valid description for local differential observables (e.g., the Frobenius norm $\|\Delta \rho_n(t)\|_F$) for all $\alpha < 1$. Indeed, the local leakage density $\langle \hat{n}_n \rangle_{\mathrm{leak}} \sim P_{\mathrm{leak}}/N \sim N^{2\alpha - 2}$ strictly vanishes as $N \to \infty$ for all $\alpha < 1$.

\end{document}